\documentclass[]{IEEETran}
\usepackage{cite}
\usepackage{amsmath,amssymb,amsfonts}
\usepackage{algorithmic}
\usepackage{graphicx}
\usepackage{textcomp}
\usepackage{xcolor}
\usepackage{subfig}
\usepackage{caption}
\usepackage{bm}
\usepackage{xspace}

\newcommand{\bmath}[1]{\ensuremath{\bm{#1}}\xspace}

\newcommand{\x}{\bmath{x}}
\newcommand{\y}{\bmath{y}}
\newcommand{\z}{\bmath{z}}

\newcommand{\f}{\bmath{f}}

\newcommand{\rv}{\bmath{r}}

\newcommand{\1}{\bmath{1}}
\newcommand{\alp}{\bmath{\alpha}}

\newcommand{\tht}{\bmath{\theta}}

\newcommand{\F}{\bmath{F}}

\newcommand{\I}{\bmath{I}}
\newcommand{\K}{\bmath{K}}

\newcommand{\Pm}{\bmath{P}}

\newcommand{\beq}{\begin{equation}}
\newcommand{\eeq}{\end{equation}}
\newcommand{\bea}{\begin{eqnarray}}
\newcommand{\eea}{\end{eqnarray}}
\newcommand{\ba}{\left(\!\!\begin{array}}
\newcommand{\ea}{\end{array}\!\!\right)}
\newcommand{\bc}{\begin{center}}
\newcommand{\ec}{\end{center}}

\newcommand{\diag}{\mathrm{diag}}

\newcommand{\Kc}{\mathcal{\K}}
\newcommand{\np}{n_\mathrm{p}}
\newcommand{\nf}{n_\mathrm{f}}
\newcommand{\nz}{n_\mathrm{z}}
\newcommand{\reb}{\mathrm{reb}}
\newcommand{\nt}{n_\mathrm{tr}}
\def\BibTeX{{\rm B\kern-.05em{\sc i\kern-.025em b}\kern-.08em
    T\kern-.1667em\lower.7ex\hbox{E}\kern-.125emX}}
\newcommand{\txtb}[1]{\textcolor{black}{#1}}

\begin{document}
	\title{{Deep Kernel Representation for Image Reconstruction in PET}}
	\author{Siqi Li and Guobao Wang
		
		\thanks{This work is supported in part by NIH under grant no. R01 DK124803. Part of this work was presented in the 2022 SPIE Medical Imaging Conference.
		}
		\thanks{S. Q. Li is with the Department of Radiology, University of California Davis Health, Sacramento, CA 95817, USA. (e-mail: sqlli@ucdavis.edu).}
		\thanks{G. B. Wang is with the Department of Radiology, University of California Davis Health, Sacramento, CA 95817, USA. (e-mail: gbwang@ucdavis.edu).}}
	
	\maketitle

	\begin{abstract}
		Image reconstruction for positron emission tomography (PET) is challenging because of the ill-conditioned tomographic problem and low counting statistics. Kernel methods address this challenge by using kernel representation to incorporate image prior information in the forward model of iterative PET image reconstruction. Existing kernel methods construct the kernels commonly using an empirical process, which may lead to \txtb{unsatisfactory} performance. In this paper, we describe the equivalence between the kernel representation and a trainable neural network model. A deep kernel method is then proposed by exploiting a deep neural network to enable automated learning of an \txtb{improved} kernel model and is directly applicable to single subjects \txtb{in dynamic PET}. The training process utilizes available image prior data to form a set of robust  kernels \txtb{in an optimized way} rather than empirically. The results from computer simulations and a real patient dataset demonstrate that the proposed deep kernel method can outperform the existing kernel method and neural network method for dynamic PET image reconstruction.
	\end{abstract}
	\begin{IEEEkeywords}
		Dynamic PET, deep kernel learning, image reconstruction, kernel methods, neural networks.
	\end{IEEEkeywords}
	\section{Introduction}
	\IEEEPARstart{P}{ositron} emission tomography (PET) is an imaging modality for quantitatively measuring biochemical and physiological processes \emph{in vivo} by using a radiotracer \cite{Vaquero2015}. Image reconstruction for PET is challenging due to the ill-conditioned tomographic problem and low-counting statistics (high noise) of PET data \cite{Qi2006}, for example, in dynamic PET imaging where short time frames are used to monitor rapid change in tracer distribution.
	
	Among different methods of PET image reconstruction, the kernel methods (e.g., \cite{Wang2015, Hutchcroft2016, Novosad2016, Wang2019,Bland2018,Deidda2019, Cheng2021}) address the noise challenge by uniquely integrating image prior information into the forward model of PET image reconstruction through a kernel representation framework \cite{Wang2015}. Image prior information may come from composite time frames of a dynamic PET scan \cite{Wang2015}, or from anatomical images (e.g., magnetic resonance (MR) images \cite{Hutchcroft2016, Novosad2016} in integrated PET/MRI). The kernel methods can be easily  implemented with the existing expectation-maximization (EM) algorithm and have demonstrated substantial image quality improvement as compared to other methods  \cite{Wang2015, Hutchcroft2016, Novosad2016}.
	
	In the existing kernel methods, a kernel representation is commonly built using an empirical process for defining feature vectors and manually selecting method-associated parameters \cite{Wang2015}. However, such an experience-based parameter tuning and optimization approach often leads to suboptimal performance. 
	In this paper, we first describe the equivalence between the kernel representation and a trainable neural network model. Based on this connection, we then propose a deep kernel method that learns the trainable components of the neural network model from available image data to enable a data-driven automated learning of an \txtb{improved} kernel model. The learned kernel model is then applied to tomographic image reconstruction and is expected to outperform existing kernel models that are empirically defined. 
	
	There are many ongoing efforts in the field to explore deep learning with neural networks for PET image reconstruction, see recent review articles, e.g.,  \cite{Reader2021, Gong2020, Wang2020, Ge2020, Ge2021, Gong2021a}. Deep neural networks have been proposed for direct mapping from the projection domain to the image domain (e.g., \cite{Haggstrom2019}) but the models are so far mainly practical for 2D data training. By unrolling an iterative tomographic reconstruction algorithm, model-based deep-learning reconstruction (e.g., \cite{Lim2020, Mehranian2020}) represents a promising direction. One limitation of this method is that it requires pre-training using a large number of data sets and involves projection data in the iterative training process, which is computationally intensive. Alternatively, neural networks can be used as ``deep image prior" for image representation in iterative reconstruction, e.g., by the convolutional neural network (CNN) model \cite{Gong2019, Gong2019a, Gong2022, Xie2020, Yokota2019, Xie2021}. It has the advantage of being directly applicable to single subjects. The resulting reconstruction problem, however, is nonlinear and is often complex and challenging to optimize. 
	
	Different from these methods that utilize pure neural networks, the proposed deep kernel method combines deep neural networks into the kernel framework \cite{Wang2015} to form a novel way for tomographic image representation. The method has a unique advantage that once the model is trained with neural networks, the unknown kernel coefficient image remains linear in the model and is therefore easy to be reconstructed from PET data. It does not necessarily require a large data set for training but is directly applicable to single-subject learning and reconstruction, \txtb{e.g., in dynamic PET}, as will be demonstrated in this paper.
	
	The rest of this paper is organized as follows. Section II introduces the background materials of the kernel method for PET image reconstruction. Section III describes the generalized theory of the proposed deep kernel method that derives a data-driven automated learning of an \txtb{improved} kernel method. We then present a computer simulation study in Section IV and a real patient data study in Section V to demonstrate the improvement of the proposed method over existing methods. Finally discussions and conclusions are drawn in Section VI and VII.
	
	\begin{figure*}[t]
		\vspace{-0pt}
		\footnotesize
		\centering
		{\includegraphics[trim=0.5cm 0cm 0cm 0cm, clip,width=5in]{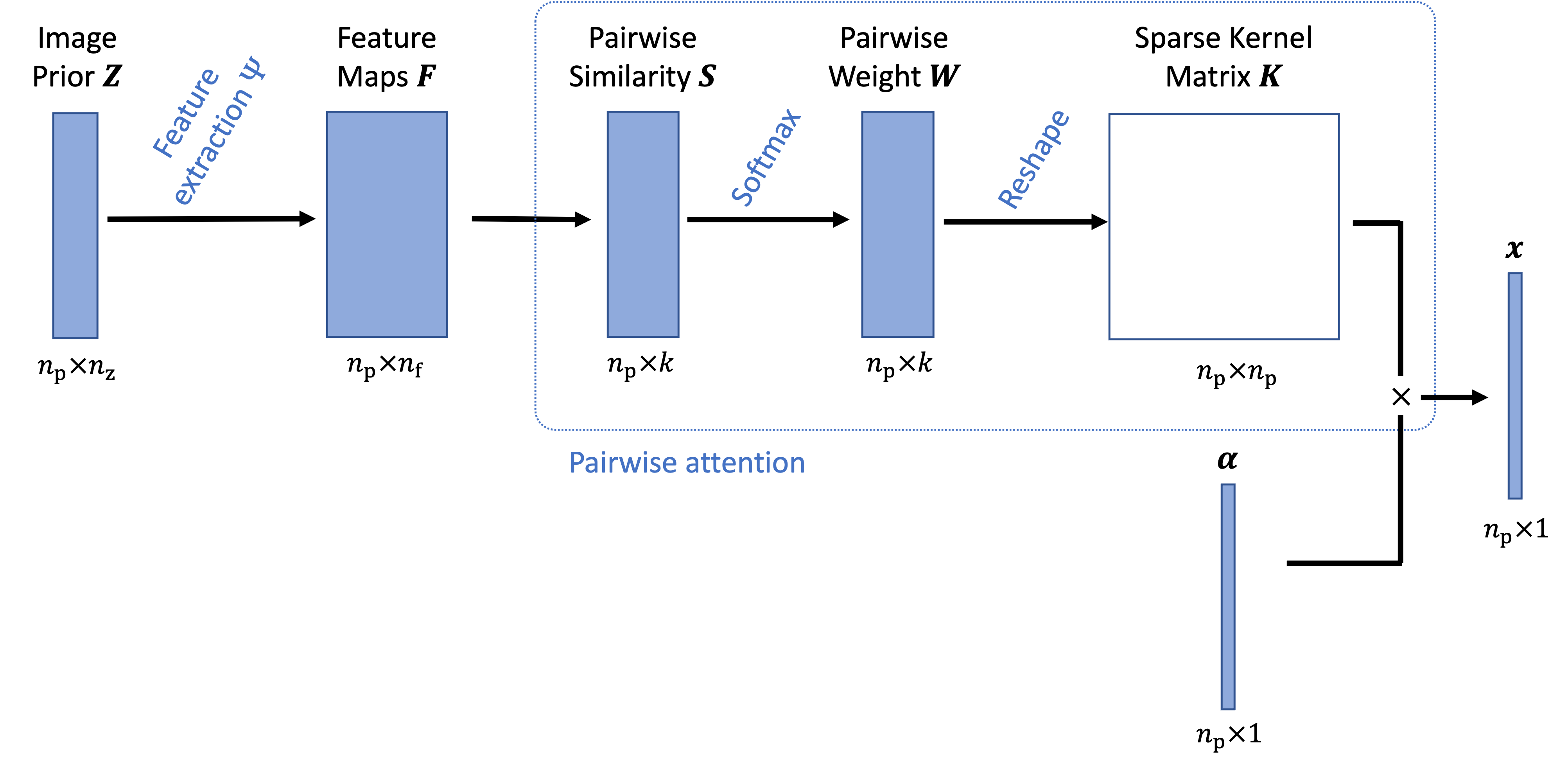}}
		\caption{The construction of kernel representation for a PET image is described as a series of neural network components.  \txtb{Both the feature extraction and pairwise attention modules are trainable.}}
		\label{fig:1}
		\vspace{-0pt}
	\end{figure*}

	\section{Background}
	\subsection{PET Image Reconstruction}
	PET projection data $\y = \left\{y_{i}\right\}$ can be well modeled as independent Poisson random variables using the log-likelihood function \cite{Qi2006},
	\beq
	L(\y|\x) = \sum_{i=1}^{N}y_{i} \log \overline{y}_{i} - \overline{y}_{i} - \log y_{i}!,
	\eeq
	where $i$ denotes the detector index and $N$ is the total number of detector pairs. The expectation of the projection data, $\overline{\y} $, is related to the unknown image $\x$ through
	\beq
	\overline{\y} = \Pm \x + \rv,
	\label{PET model}
	\eeq
	where $\Pm $ is the detection probability matrix for PET and includes normalization factors for scanner sensitivity, scan duration, deadtime correction and attenuation correction. $\rv$ is the expectation of random and scattered events \cite{Qi2006}.
	
	The maximum likelihood (ML) estimate of the image $\x$ is found by maximizing the Poisson log-likelihood,
	\beq
	\hat{\x} =\arg\max\limits_{\x \ge 0}L(\y|\x).
	\label{ML}
	\eeq
	A common way of seeking the solution of (\ref{ML}) is to use the EM algorithm \cite{Shepp1982}.
	
	\subsection{Kernel Methods for PET Reconstruction}
	
	The kernel methods describe the image intensity $x_j$ at the pixel $j$ as a linear representation of kernels \cite{Wang2015},
	\beq
	x_j = \sum_{l\in \mathcal{N}_j} \alpha_l \kappa(\f_j,\f_l),\quad j,l=1,\cdots,\np
	\label{kernel representation}
	\eeq
	where $\mathcal{N}_j$ defines the neighborhood of pixel $j$ and $\np$ is the total number of image pixels. $\f_j$ and $\f_l$ are the feature vectors extracted from image priors for pixel $j$ and pixel $l$, respectively.  $\alpha_l$ is the kernel coefficient at pixel $l$. $\kappa(\cdot,\cdot)$ is the kernel function that defines a weight between pixel $j$ and pixel $l$. A popular choice of $\kappa(\cdot,\cdot)$ is the radial Gaussian kernel,
	\beq
	\kappa(\f_j,\f_l) = \mathrm{exp}\big(-\frac{||\f_j-\f_l||^2}{2\sigma^2}\big),
	\label{Gaussian}
	\eeq
	with $\sigma$ being the kernel parameter.
	The equivalent matrix-vector form of (\ref{kernel representation}) is
	\beq
	\x = \K \alp
	\label{eq-km}
	\eeq
	with the $(j,l)$th element of the square kernel matrix $\K$ being $\kappa(\f_j,\f_l)$. 
	
	The kernel coefficient image $\alp$ is then estimated from the projection data $\y$ by maximizing the log-likelihood $L$,
	\beq
	\hat{\alp} =\arg\max\limits_{\alp \ge 0}L(\y|\K\alp),
	\label{ML-K}
	\eeq
	which can be solved using the kernelized EM algorithm \cite{Wang2015},
	\beq
	\alp^{n+1} = \frac{\alp^n}{\K^T\Pm^T\1_N}\cdot\left( \K^T\Pm^T\frac{\y}{\Pm\K\alp^n+\rv}\right),
	\label{KEM}
	\eeq
	where $n$ denotes the iteration number and the superscript ``$T$'' denotes matrix transpose. $\1_{N}$ is a vector with all elements being 1.
	Once $\alp$ is estimated, the final PET activity image $\x$ is given by
	$
	\hat{\x} = \K\hat{\alp}.
	$
	
	Note that in practice, a normalized kernel matrix
	\beq
	\bar{\K}=\diag^{-1}[\K\1_N]\K
	\label{eq-normk}
	\eeq
	is commonly used for better performance \cite{Wang2015}. The ($j$,$l$)th element of $\bar{\K}$ is equivalent to
	\beq
	\kappa(\f_j,\f_l) = 
	\frac{\mathrm{exp}\big(-\frac{||\f_j-\f_l||^2}{2\sigma^2}\big)}{\sum_{l' \in \mathcal{N}_j}\mathrm{exp}\big(-\frac{||\f_j-\f_{l'}||^2}{2\sigma^2}\big)},\quad l\in \mathcal{N}_j .
	\label{Gaussian}
	\eeq
	The neighborhood $\mathcal{N}_j$ of pixel $j$ can be defined by its $k$-nearest neighbors (kNN) \cite{Friedman1977} to make $\K$ sparse. The feature vector $\f_j$ is usually set to the intensity values of the image prior at pixel $j$ and the kernel parameter $\sigma$ is chosen empirically, e.g. $\sigma=1.0$.

	\section{Proposed Deep Kernel Method}
	
	\subsection{Kernel Representation as Neural Networks}
	
	We first describe the kernel representation using a neural network description  illustrated in Fig. \ref{fig:1}. The construction of kernel representation is decomposed into two main modules: (1) feature extraction and (2) pairwise attention.
	
	Denote the image prior data by $\bmath{Z}$ which consists of  $\nz$ prior images, \txtb{each with $\np$ pixels.} The feature extraction module is to extract a feature vector $\f$ of length $\nf$ for each image pixel from $\bmath{Z}$,
	\beq
	\f_j = \Psi_j(\bmath{Z}),
	\eeq
	where $\Psi$ denotes the feature extraction operator, for example, a convolutional neural network. The extraction of conventional intensity-based features is equivalent to a $1\times 1\times 1$ convolution operation on $\bmath{Z}$ (if the images are 3D). This step provides a feature data $\F$ of size $\np\times\nf$. 
	
	The pairwise attention module first calculates the similarity between pixel $j$ and its neighboring pixels that are specified by a pre-determined neighborhood $\mathcal{N}_j$, 
	\beq
	s_{jl}=-\frac{||\f_j-\f_l||^2}{2\sigma^2}, l\in \mathcal{N}_j.
	\label{similarity}
	\eeq
	Note that here only  $k$ neighbors are selected for each pixel $j$, \txtb{e.g., using the Euclidean distance-based kNN algorithm \cite{Wang2015, Friedman1977}}. This leads to a similarity data $\bmath{S}$ of size $\np\times k$. Pairwise weights are then calculated from $\bmath{S}$ using 
	\beq
	w_{jl}= \frac{\mathrm{exp}(s_{jl})}{\sum_{l' \in \mathcal{N}_j}\mathrm{exp}(s_{jl'})},
	\label{softmax}
	\eeq 
	generating a weight data $\bmath{W}$ of size $\np\times k$. Here $\sum_{l\in \mathcal{N}_j} w_{jl}=1$. This type of weight calculation is also called \textit{softmax} in neural networks and can be directly explained as a pairwise attention mechanism \cite{Vaswani2017, Wang2018}. $w_{jl}$ is the attention weight of other ``key" pixels $\{l\}$ as compared to the ``query" pixel $j$.
	
	The final step reshapes $\bmath{W}$ using the neighborhood indices defined by $\{\mathcal{N}_j\}_{j=1}^{\np}$ to generate a sparse matrix, which is equal to the normalized kernel matrix defined in (\ref{eq-normk}). Each row of the kernel matrix is of the size $\np\times 1$ and can be displayed as an image, which can also be understood as an attention map for the corresponding pixel in $\alp$. 
	
	\subsection{Deep Kernel Model}
	
	Integrating all the neural network components in Fig. \ref{fig:1} together, we have the following deep kernel model to represent a PET image $\x$,
	\beq
	\x=\Kc(\tht;\bmath{Z})\alp,
	\eeq
	where $\Kc(\tht;\bmath{Z})$ denotes the equivalent neural network model of $\K$ with the image prior data $\bmath{Z}$ as the input and $\tht$ collecting any model parameters that are trainable. 
	
	The deep kernel model is nonlinear with respect to $\tht$ and $\bmath{Z}$ but remains linear with respect to the kernel coefficient image $\alp$. While this model shares the spirit of using attention with the nonlocal neural network \cite{Wang2018}, the linearity of $\alp$ makes it unique and more suitable for tomographic image reconstruction problems. Once $\tht$ is determined, $\alp$ can be easily reconstructed from the projection data $\y$ using the kernelized EM algorithm in (\ref{KEM}).
	
	In conventional kernel methods, $\tht$ is equivalent to be determined empirically, which \txtb{does not explore the full potential of the kernel method}. For example, intensity-based features are commonly used for $\f$. However, convolutional neural network-derived features can be more informative \cite{Li2021}. In this paper, we exploit the capability of deep learning to train an optimized feature set for generating $\K$ from available image prior data based on the proposed deep kernel model.
	
	\subsection{Deep Kernel Learning}
	
	The deep kernel learning problem is formulated using the observation that in the kernelized image model  (\ref{eq-km}), $\x$ is usually a clean version of $\alp$  if $\alp$ is noisy. This inspires the following use of the denoising autoencoder framework \cite{Kramer1991}  to train the model parameters of $\Kc(\tht;\bmath{Z})$,
	\beq
	\hat{\tht} = \arg\min\limits_{\tht}\sum_{q=1}^{\nt}||\I_q - \Kc(\tht;\bmath{Z})\tilde{\I}_q||^2,
	\label{supervised}
	\eeq
	where $\I_q$ denotes the $q$th high-quality image in the training dataset and $\tilde{\I}_q$ is a corrupted version of $\I_q$. $\nt$ is the total number of training image pairs. In PET, $\I_q$  and $\tilde{\I}_q$ can be obtained from high count data and low-count data, respectively.
	
	The deep kernel model can be pretrained using a large number of patient scans (large $\nt$) if such a training dataset is available. It can also be trained online for single subjects (small $\nt$) without pretraining, as described below.

	\subsection{Single-Subject Deep Kernel Method for Dynamic PET}
	
	In dynamic PET, the image prior data $\bmath{Z}$ may consist of several composite images $\{\z_m\}_{m=1}^{\nz}$ where $\nz$ is the number of composite frames. These images are reconstructed from the rebinned long-scan projection data $\{\y_m^{\reb}\}_{m=1}^{\nz}$ and may have good image quality due to the relatively high count level of a composite frame. \txtb{For example, a one-hour dynamic $^{18}$F-fluorodeoxyglucose (FDG) PET scan can be divided into three composite frames, each of 20 minutes \cite{Wang2015}.} The composite image prior has been used in the standard kernel methods for constructing the kernel matrix empirically. Here we use it to train an \txtb{improved} kernel model adaptive to a single subject.
	
	The single-subject deep kernel learning problem \txtb{for dynamic PET} is \txtb{constructed using the following optimization criterion},
	\beq
	\hat{\tht} = \arg\min\limits_{\tht}\sum_{m=1}^{\nz}||\z_m - \Kc(\tht;\bmath{Z})\tilde{\z}_m||^2,
	\label{supervised}
	\eeq
	where the corrupted image $\tilde{\z}_m$ can be obtained from the reconstruction of \txtb{the low-count projection data which are downsampled from $\y_m^{\reb}$} using a count reduction factor $d$ (e.g. $d=10$). Once $\tht$ is trained, the learned kernel model is then used to reconstruct all the dynamic frames of the scan frame-by-frame using the kernel EM algorithm in  (\ref{KEM}).
	
	In theory, both the feature extraction and pairwise attention modules in the neural network model (Fig. \ref{fig:1}) are trainable. As a proof of concept, in this work we only train the feature extraction operator $\Psi$, \txtb{while the pairwise attention module is calculated using (\ref{similarity}) and (\ref{softmax}) as used in the conventional kernel method \cite{Wang2015}.} 
	
	\subsection{Model Structure of Feature Extraction}
	
\begin{figure}[t]
		\centering
		\includegraphics[trim=2cm 0cm 2cm 0cm, width=3in]{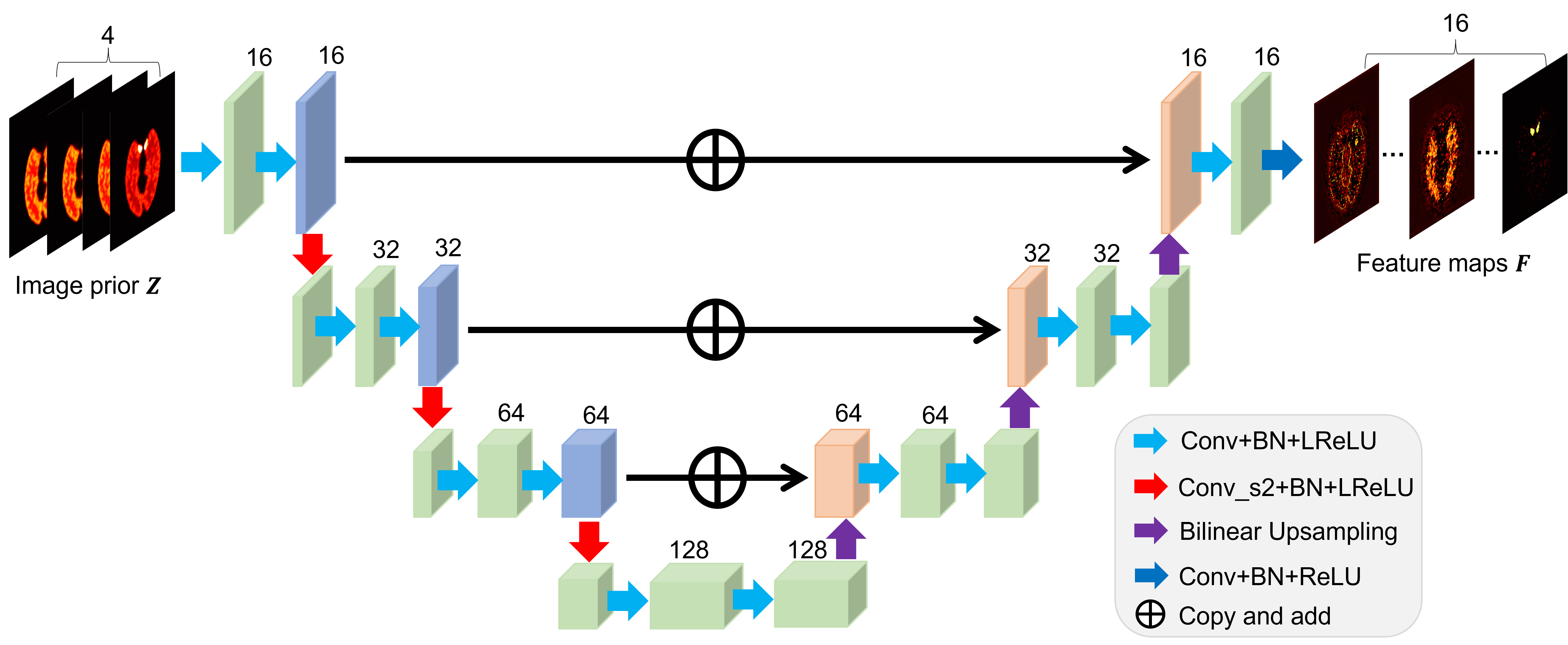}
		\caption{Illustration of a residual U-net $\Psi$ used for feature extraction in this work.}
		\label{fig:Unet}	
	\end{figure}

	\txtb{The proposed method is applicable to different neural network architectures if they are suitable for image representation.} Here a \txtb{popular}  residual U-net architecture \cite{Gong2019}, as illustrated in Fig. \ref{fig:Unet}, is used for the feature extraction module $\Psi$. The network is available in both 2D and 3D versions for learning 2D and 3D images, respectively. It consists of the following operations: 1) 3$\times$3 ($\times$3) 2D (3D) convolutional layer, 2) 2D (3D) batch normalization (BN) layer, 3) leaky rectified linear unit (LReLU) layer, 4) 3$\times$3 ($\times$3) convolutional layer with stride 2$\times$2 ($\times$2) for down-sampling, 5) 2$\times$2 ($\times$2) bilinear (trilinear) interpolation layer for up-sampling, 6) identity mapping layer that adds feature maps from left-side encoder path to the right-side decoder path. In addition, a ReLU layer is used before the output in order to satisfy the non-negative constraint on the last feature map.

	\section{Computer Simulation Validation}
	
	\subsection{Simulation Setup}
	
\begin{figure}[t]
		\centering
		\subfloat[]{\includegraphics[trim=0cm 0cm 0cm 1cm, width=2in]{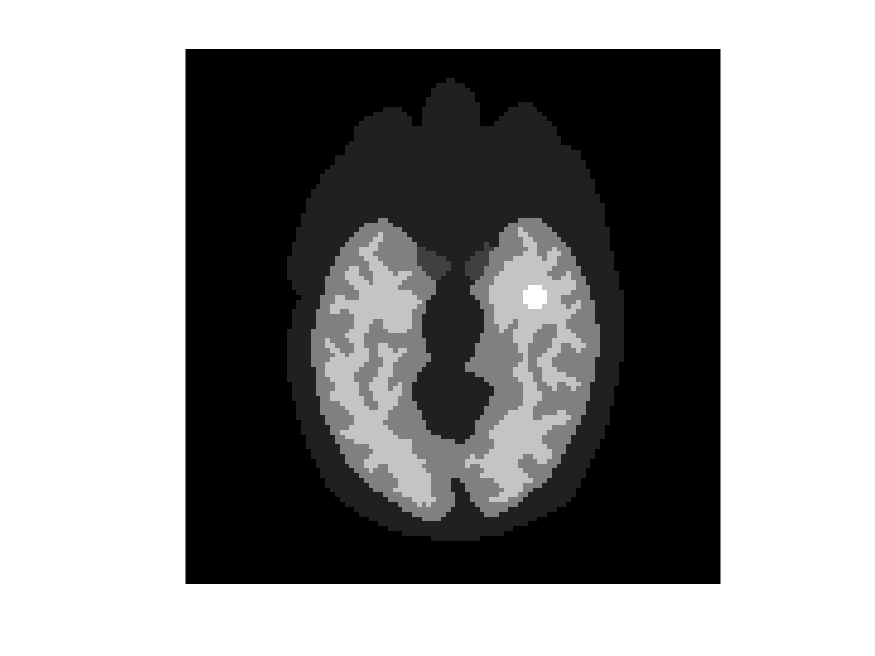}}\\
		\subfloat[]{\includegraphics[trim=0cm 0cm 0cm 1cm, width=1.8in]{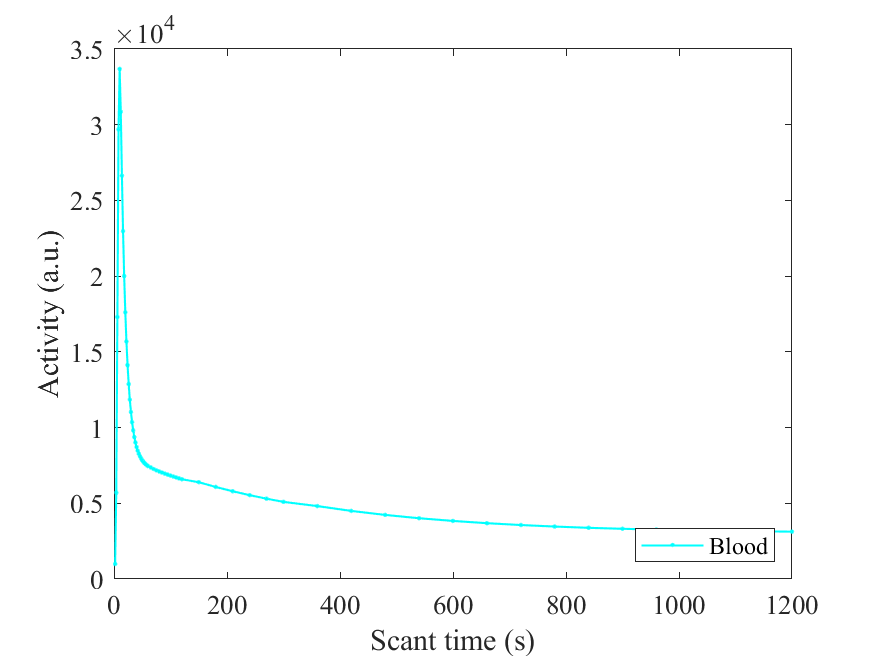}}
		\subfloat[]{\includegraphics[trim=0cm 0cm 0cm 0cm, width=1.8in]{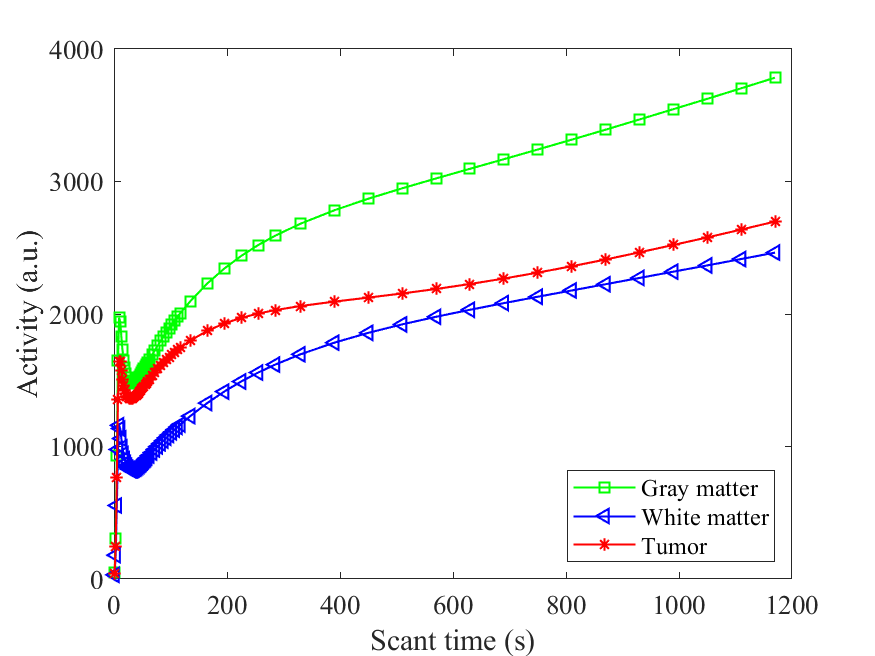}}
		\caption{Digital phantom and time activity curves used in the simulation study. (a) Zubal brain phantom composed of gray matter, white matter, blood pools and a tumor; (b) blood input function; (c) Regional time activity curves of brain regions.}
		\label{fig:phant}	
	\end{figure}

	Dynamic $^{18}$F-FDG PET scans were simulated for a GE DST whole-body PET scanner in two-dimensional mode using a Zubal head phantom shown in Fig. \ref{fig:phant}a. The phantom is composed of gray matter, white matter, blood pools and a tumor (15 mm in diameter).  A early 20-minute dynamic scan was divided into 63 time frames: 30$\times$2s, 12$\times$5s, 6$\times$30s, and 15$\times$60s. The pixel size is 3$\times$3 mm$^2$ and the image size is 111$\times$111.
	
	The time activity curves of different regions are shown in Fig. \ref{fig:phant}(b-c). An attenuation map was simulated with a constant linear attenuation coefficient assigned in the whole brain.  Dynamic images were first forward projected to generate noise-free sinograms. Poisson noise was then introduced. 
	A 20\% uniform background was included to account for mean random and scatter events. The expected total number of events over 20 min was 20 million. Twenty noisy realizations were simulated and each was reconstructed independently for comparison.
	
	\begin{figure*}[t]
		\vspace{-30pt}
		\centering
		\subfloat[]{
			\begin{minipage}[t]{0.22\linewidth}
				\centering
				\includegraphics[trim=3.3cm 0.9cm 2.3cm 0cm, clip, height=4cm]{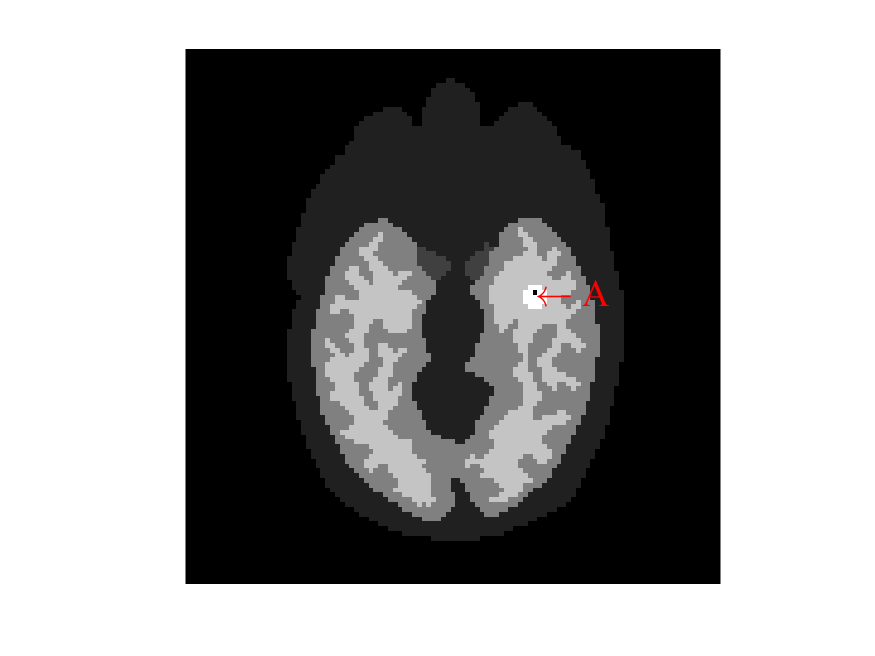}\\
				\includegraphics[trim=3.3cm 0.9cm 2.3cm 0cm, clip, height=4cm]{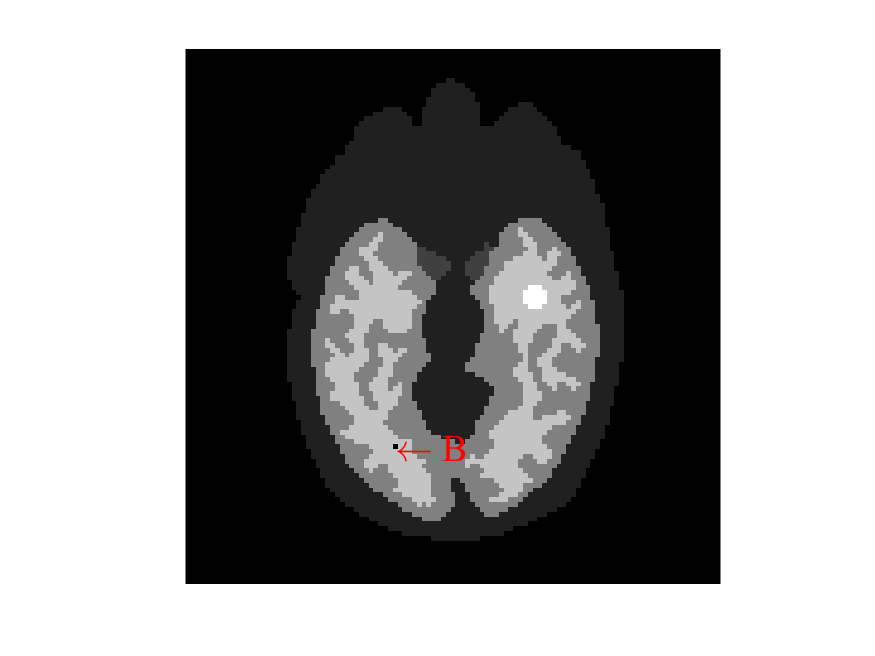}
			\end{minipage}%
		}
		\subfloat[]{
			\begin{minipage}[t]{0.22\linewidth}
				\centering
				\includegraphics[trim=3.4cm 1cm 3.4cm 0.5cm, clip, height=4cm]{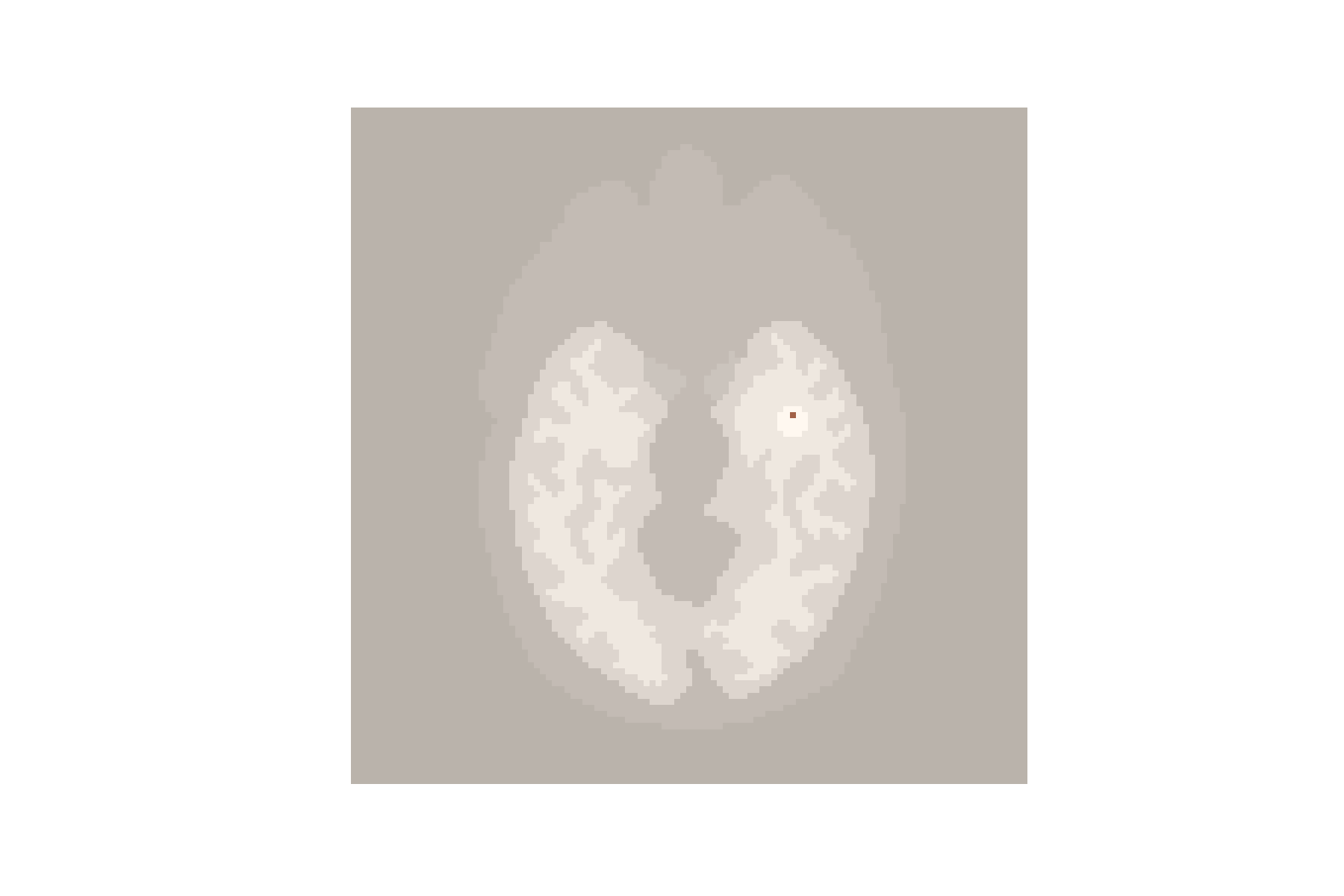}\\
				\includegraphics[trim=3.4cm 1cm 3.4cm 0.5cm, clip, height=4cm]{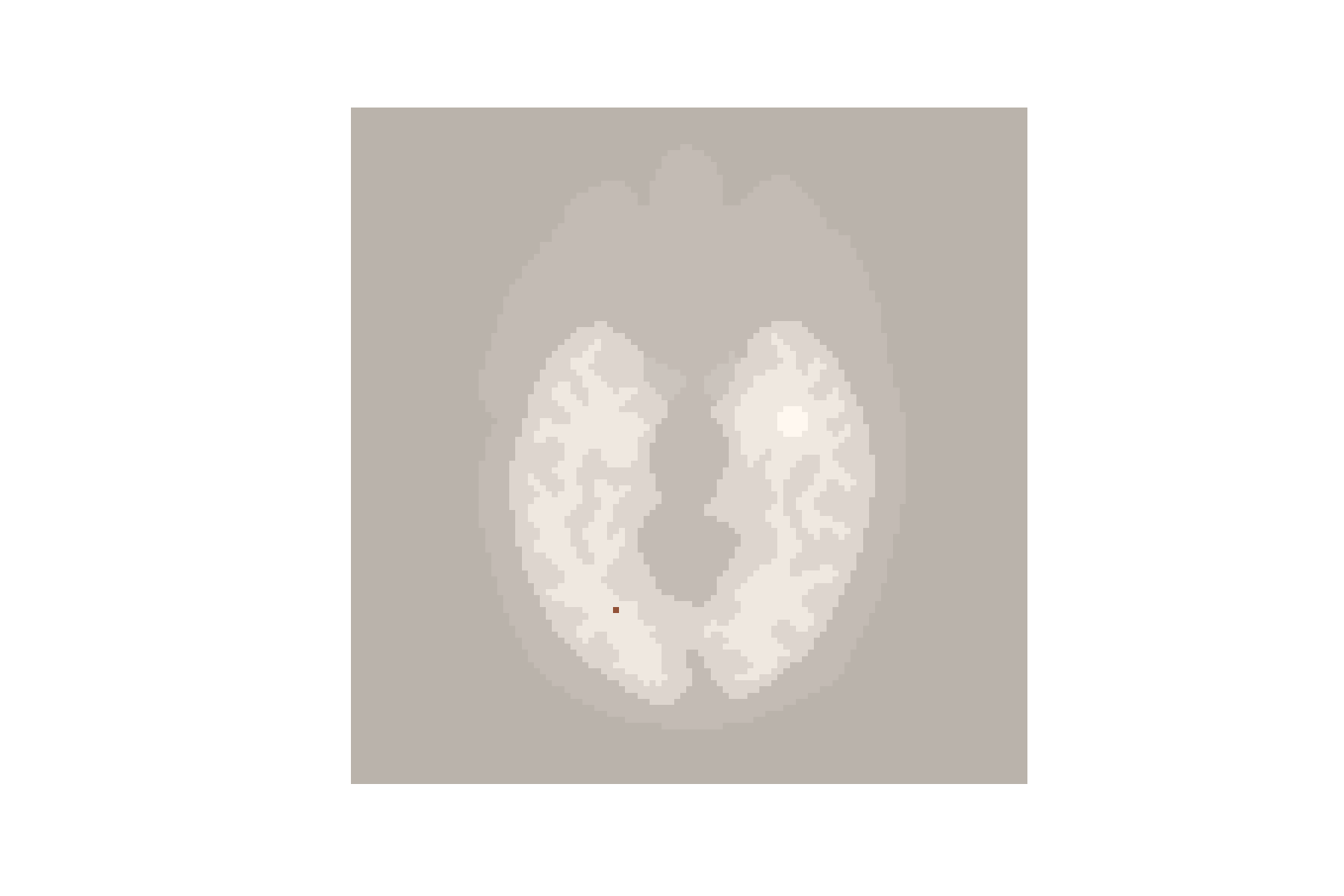}
			\end{minipage}%
		}
		\subfloat[]{
			\begin{minipage}[t]{0.22\linewidth}
				\centering
				\includegraphics[trim=3.4cm 1cm 3.4cm 0.5cm, clip, height=4cm]{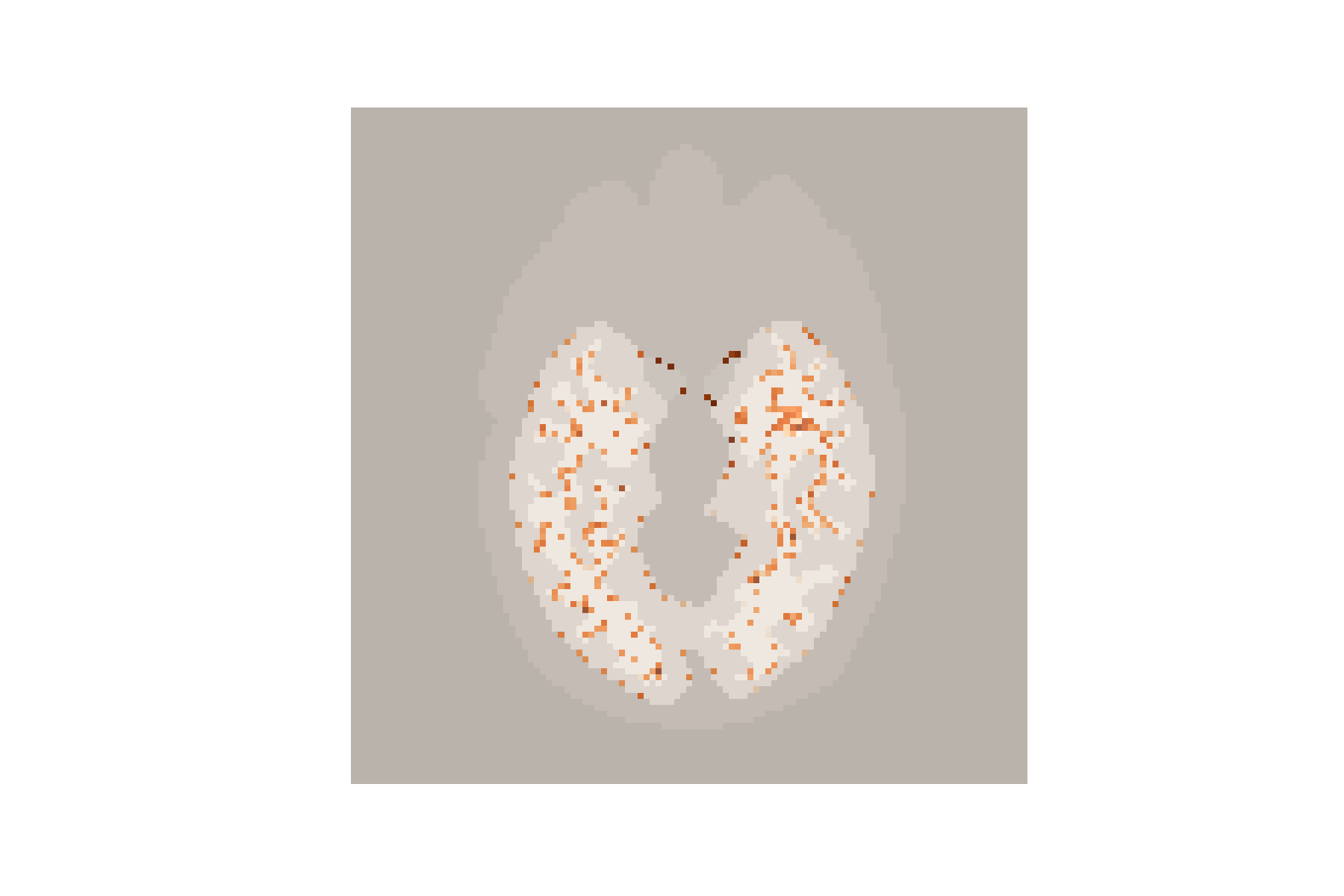}\\
				\includegraphics[trim=3.4cm 1cm 3.4cm 0.5cm, clip, height=4cm]{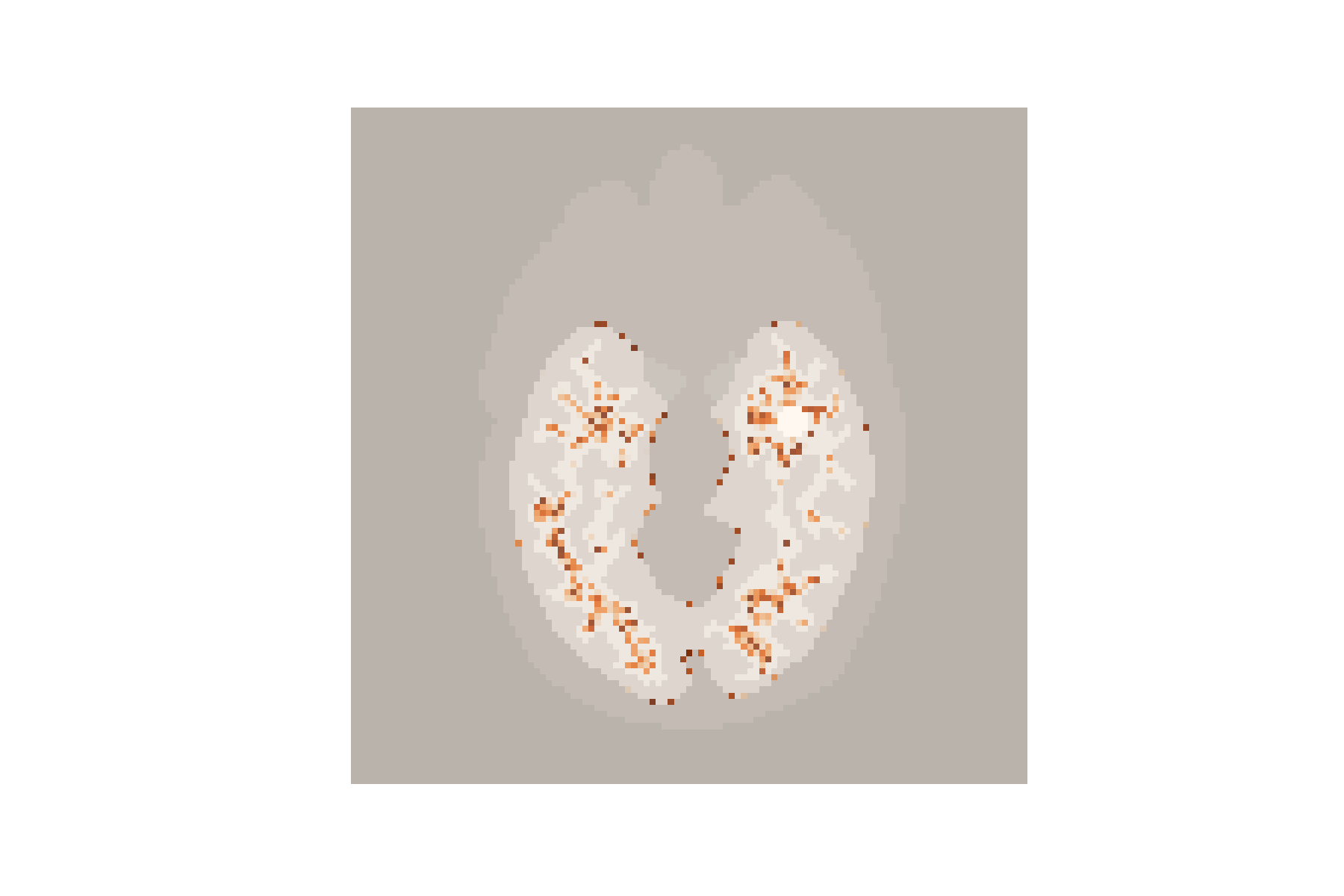}
			\end{minipage}%
		}
		\subfloat[]{
			\begin{minipage}[t]{0.22\linewidth}
				\centering
				\includegraphics[trim=3.4cm 1cm 3.4cm 0.5cm, clip, height=4cm]{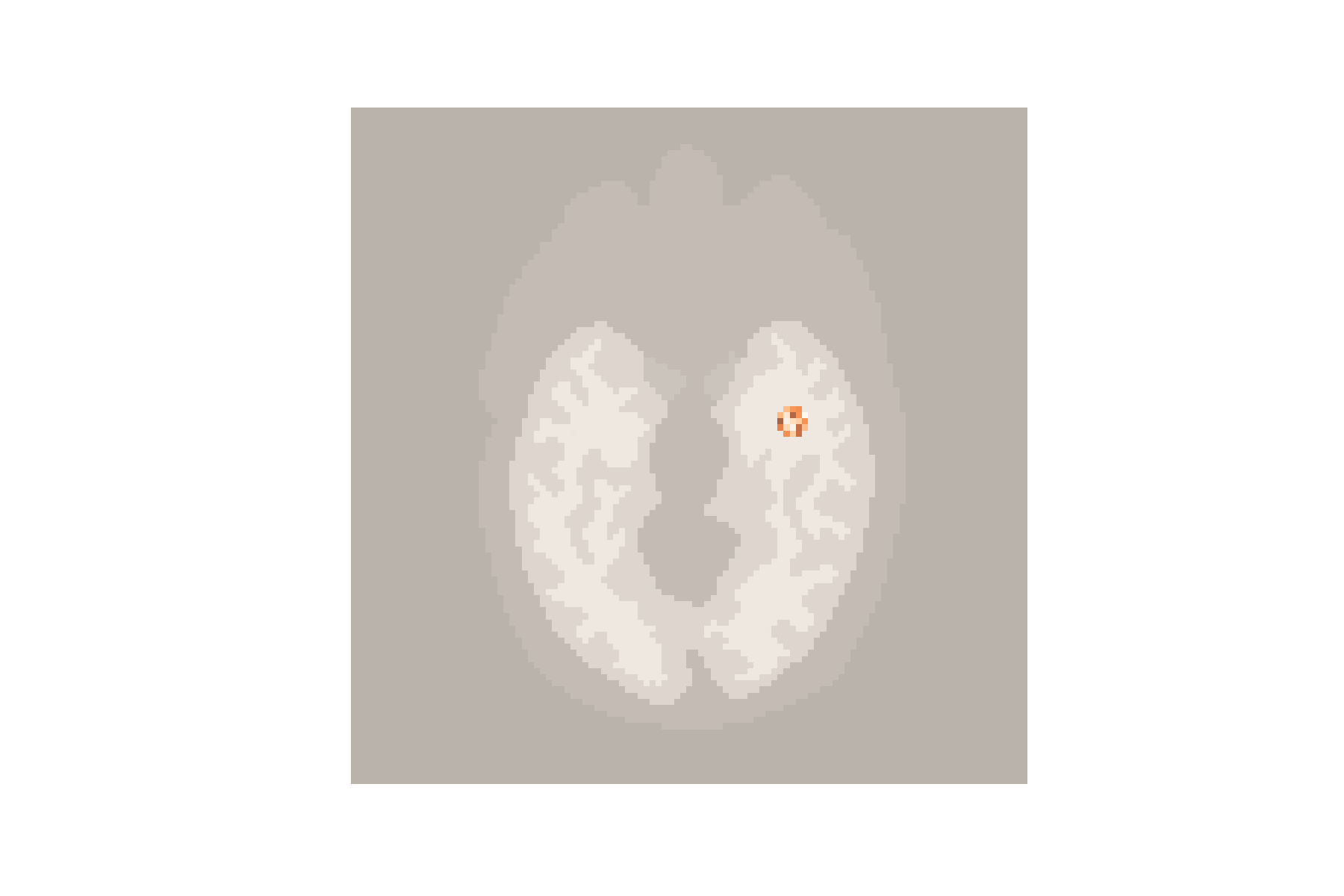}\\
				\includegraphics[trim=3.4cm 1cm 3.4cm 0.5cm, clip, height=4cm]{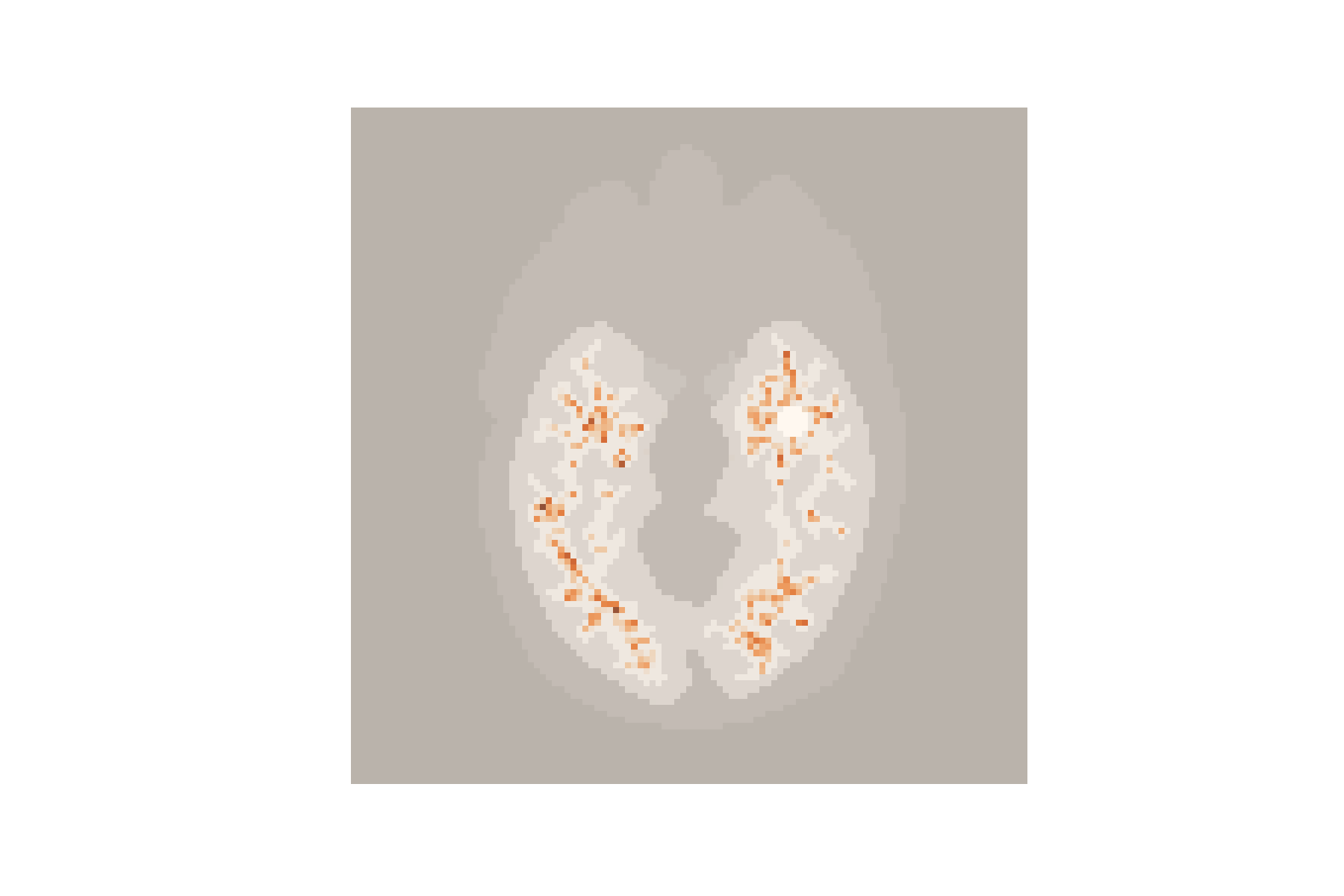}
			\end{minipage}%
		}%
		\caption{Illustration of two query pixels and the attention maps provided by different methods. (a) location of the two query pixels - one in the tumor region (top, A) and the other in the white matter (bottom, B),  (b-c) attention maps by traditional ML-EM (b), conventional kernel method (c), and the proposed deep kernel method (d). All the attention maps are overlaid on the structural image.}
		\label{fig:attention}
	\end{figure*}
	\subsection{Reconstruction Methods}
	
	The simuated dynamic data were reconstructed using four different methods: (1) standard ML-EM reconstruction; (2) existing kernel EM \cite{Wang2015}; (3) the  deep image prior (DIP) reconstruction method \cite{Gong2019} as a recent representative of nonlinear neural network-based reconstruction methods; and (4) proposed deep kernel method with single-subject online training of the feature extraction module. \txtb{The deep kernel method was trained separately for each of the 20 noisy data realizations.} All reconstructions were run for 200 iterations with a uniform initial image.
	
	The image priors for the kernel methods were the composite images obtained from four composite frames, each with 5 min scan. For the conventional kernel method, pixel intensity values extracted from the composite images {\txtb{$\{\z_m\}$}} were used to form the feature vector $\f$ for generating the kernel matrix $\K$ using kNN with $k=48$ in the same way as used in \cite{Wang2015}. 
	
	The DIP method was implemented using the alternating direction method of multipliers (ADMM) in a way similar to \cite{Gong2019} but was adapted to use the composite image prior data as the input of the U-net. Within each outer iteration, 4 iterations were used for solving the penalized-likelihood image reconstruction problem and 50 iterations were used for the image-domain DIP learning. These settings were empirically optimized for obtaining stable results \txtb{across different time frames} according to the image mean squared error (MSE)  in our experiments. The effect of the ADMM hyper-parameter $\rho$ was also investigated and $ \rho= 5\times10^{-6}$ was chosen to obtain nearly optimal image MSE.
	
	In the deep kernel method, the low-count images $\{\tilde{\z}_m\}$ were obtained by using one-tenth of the counts in each composite frame $\z_m$. \txtb{ML-EM was used to reconstruct the image pair $\{\tilde{\z}_m\}$ and $\{\z_m\}$ for training}. The $k$ in kNN for defining the neighborhood $\{\mathcal{N}_j\}$ was set to be 200 for optimized image MSE performance. For implementation, the tomographic reconstruction step was implemented in MATLAB and the deep kernel training step was implemented in PyTorch, both on a PC with an Intel i9-9920X CPU with 64GB RAM and a NVIDIA GeForce RTX 2080Ti GPU. Three hundred iterations were used for the training step with the learning rate set to $10^{-3}$. \txtb{The Kaiming initialization method\cite{He2015} was used for each convolutional layer and uniform initialization was used for each BN layer.}
	
	\subsection{Evaluation Metrics}
	
	Different image reconstruction methods were compared using the image MSE defined by
	\beq
	{\rm MSE}(\hat{\x}_m) = 10\log_{10}\big( ||\hat{\x}_m - \x_m^{\rm{true}}||^2/||\x_m^{\rm{true}}||^2\big) (\mathrm{dB}),
	\eeq
	where $\hat{\x}_m$ is an image estimate of frame $m$ obtained with one of the reconstruction methods and $\x_m^{\rm{true}}$ denotes the ground truth image. The ensemble bias and standard deviation (SD) of the mean intensity in regions of
	interest (ROIs) were also calculated to evaluate ROI quantification,
	\beq
	\rm{Bias} = \frac{1}{\rm{c}^{\rm{true}}}\big|\overline{\rm{c}} - \rm{c}^{\rm{true}}\big|,\;
	{\rm SD} =\frac{1}{\rm{c}^{\rm{true}}}\sqrt{\frac{1}{N_r-1}\sum_{i=1}^{N_r}|\rm{c}_{\it{i}}-\overline{\rm{c}}|},
	\eeq
	where $\rm{c}^{\rm{true}}$ is the noise-free intensity and $\overline{\rm{c}}=\frac{1}{N_r}\sum_{i=1}^{N_r}\rm{c}_{\it{i}}$ denotes the mean of $N_r$ realizations. $c_i$ is the mean ROI uptake in the $i$th realization. 

	\begin{figure*}[t]
		\vspace{-20pt}
		\centering
		\subfloat[]{
			\begin{minipage}[t]{0.22\linewidth}
				\centering
				\includegraphics[trim=2cm 1cm 0.8cm 0cm, clip, height=3.2cm]{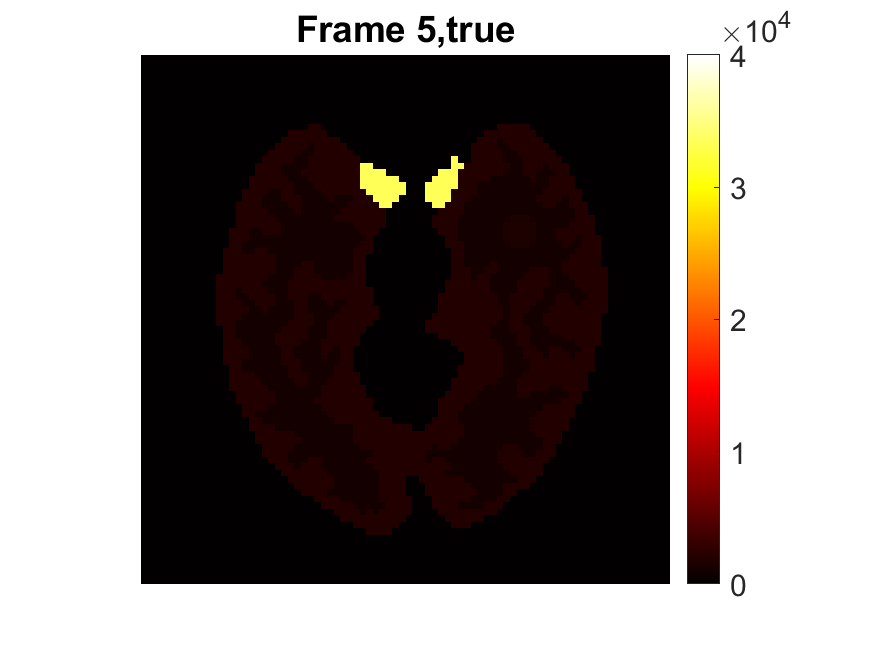}\\
				\includegraphics[trim=2cm 1cm 0.8cm 0cm, clip, height=3.2cm]{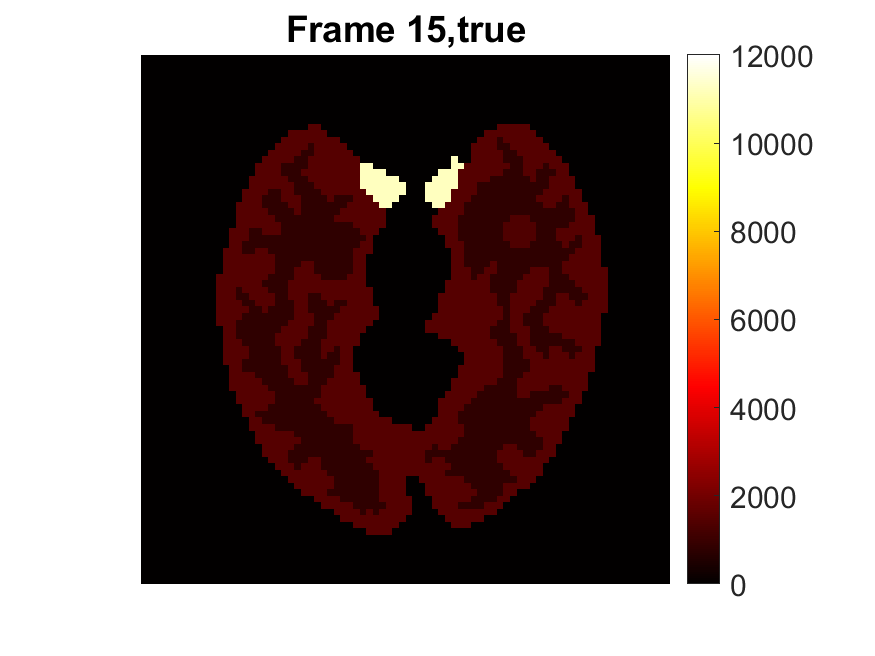}\\
				\includegraphics[trim=2cm 1cm 0.8cm 0cm, clip, height=3.2cm]{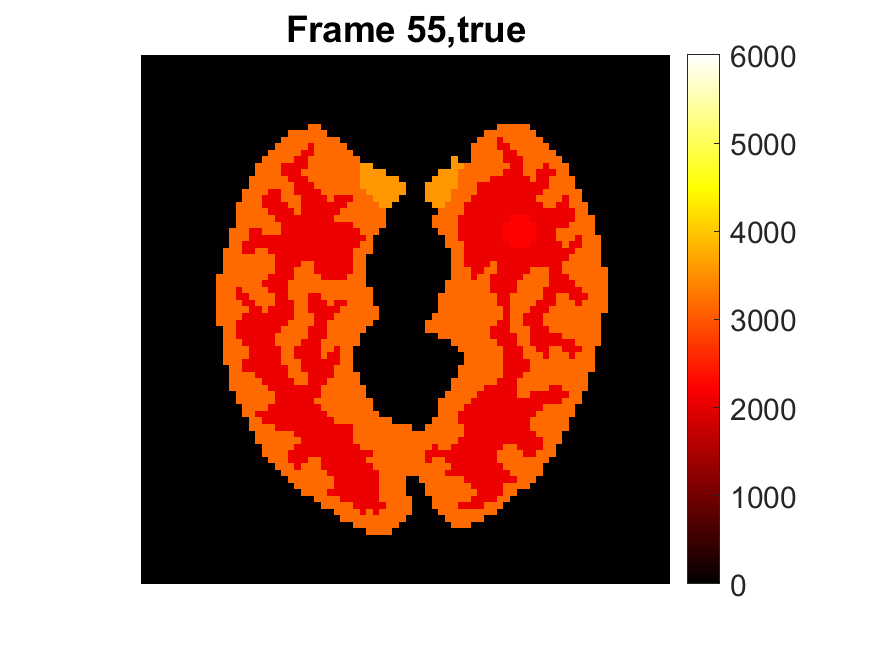}\\
			\end{minipage}%
		}%
		\subfloat[]{
			\begin{minipage}[t]{0.18\linewidth}
				\centering
				\includegraphics[trim=2cm 1cm 3.4cm 0cm, clip, height=3.2cm]{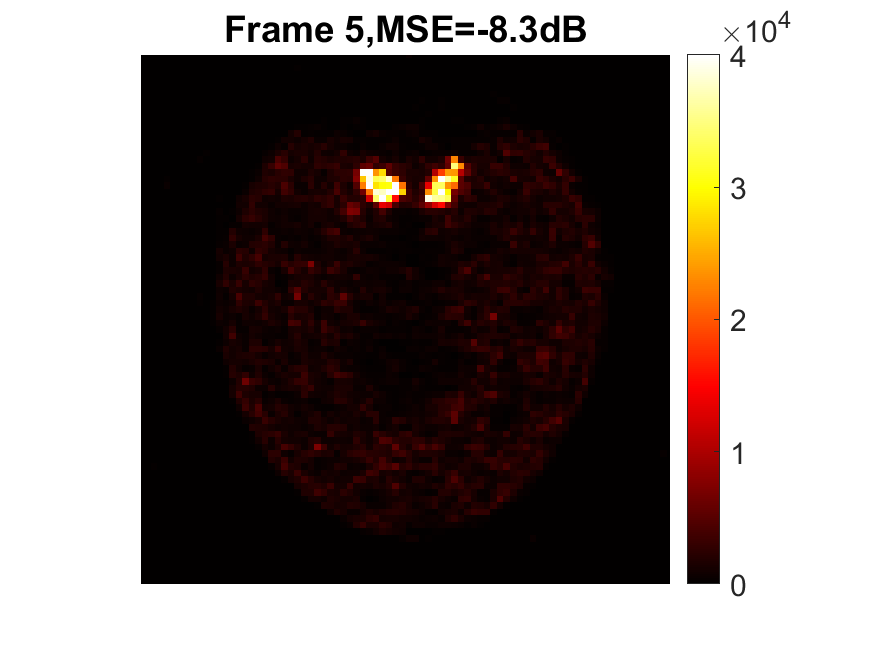}\\
				\includegraphics[trim=2cm 1cm 3.4cm 0cm, clip, height=3.2cm]{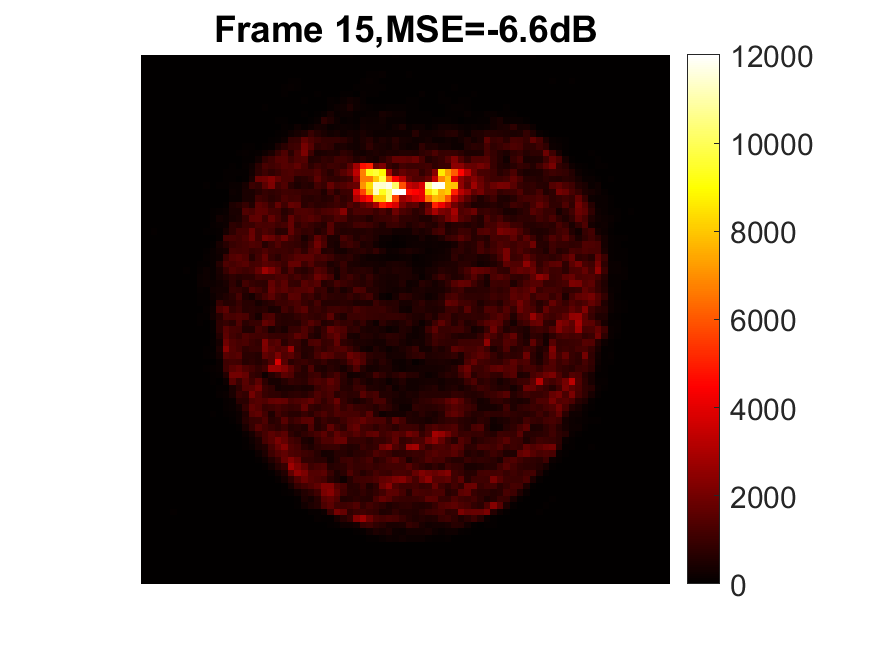}\\
				\includegraphics[trim=2cm 1cm 3.4cm 0cm, clip, height=3.2cm]{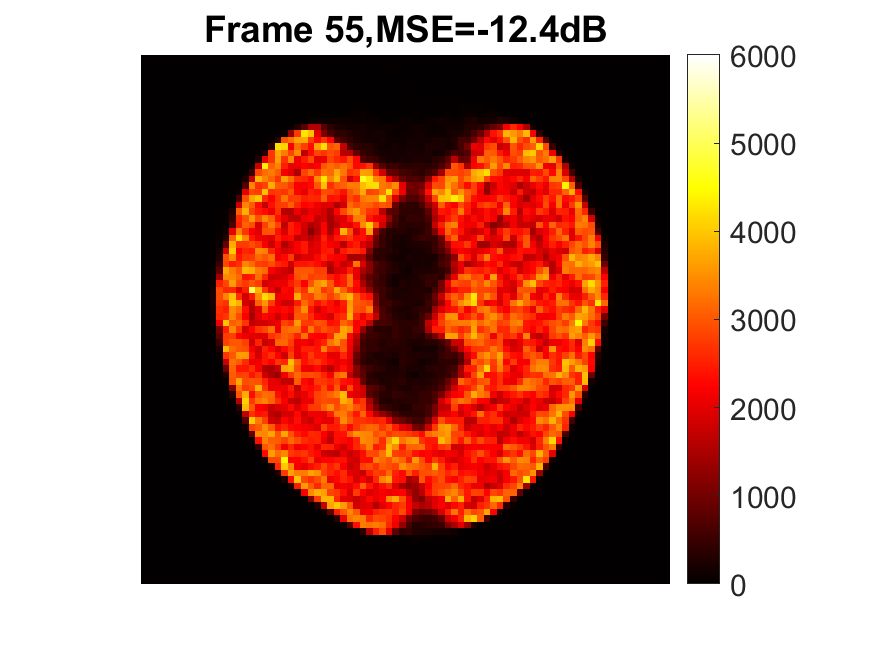}\\
			\end{minipage}%
		}%
		\subfloat[]{
			\begin{minipage}[t]{0.18\linewidth}
				\centering
				\includegraphics[trim=2cm 1cm 3.4cm 0cm, clip, height=3.2cm]{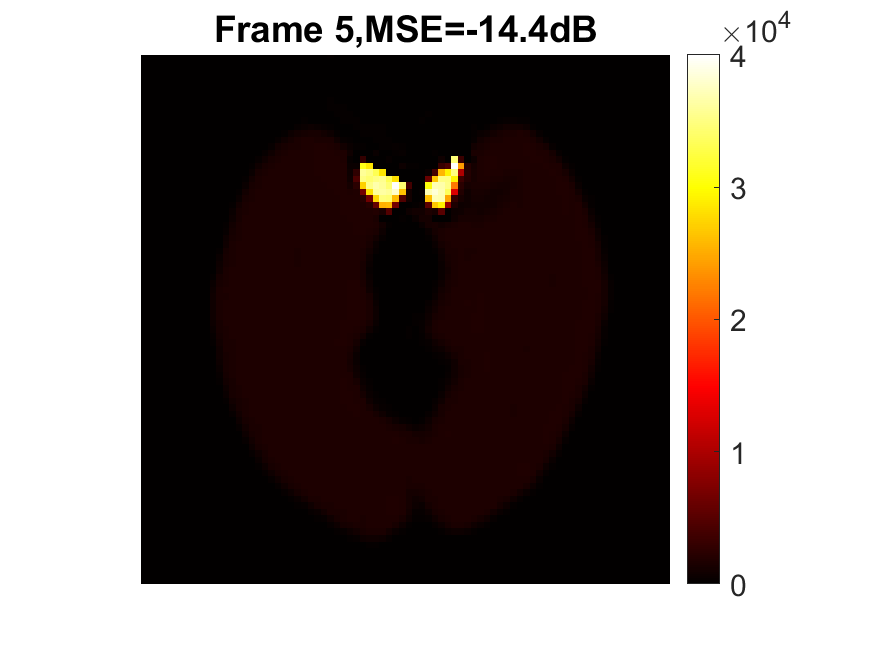}\\
				\includegraphics[trim=2cm 1cm 3.4cm 0cm, clip, height=3.2cm]{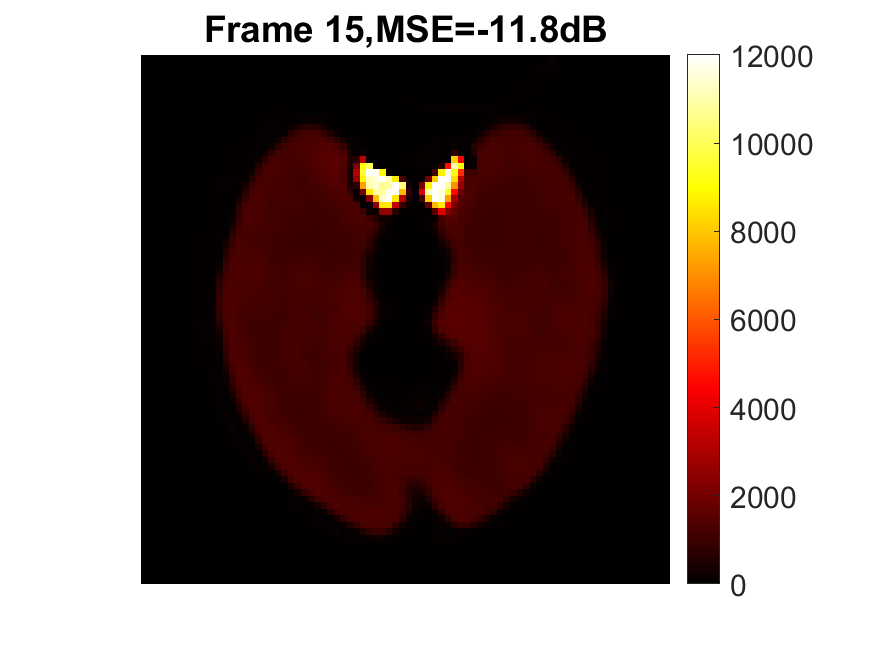}\\
				\includegraphics[trim=2cm 1cm 3.4cm 0cm, clip, height=3.2cm]{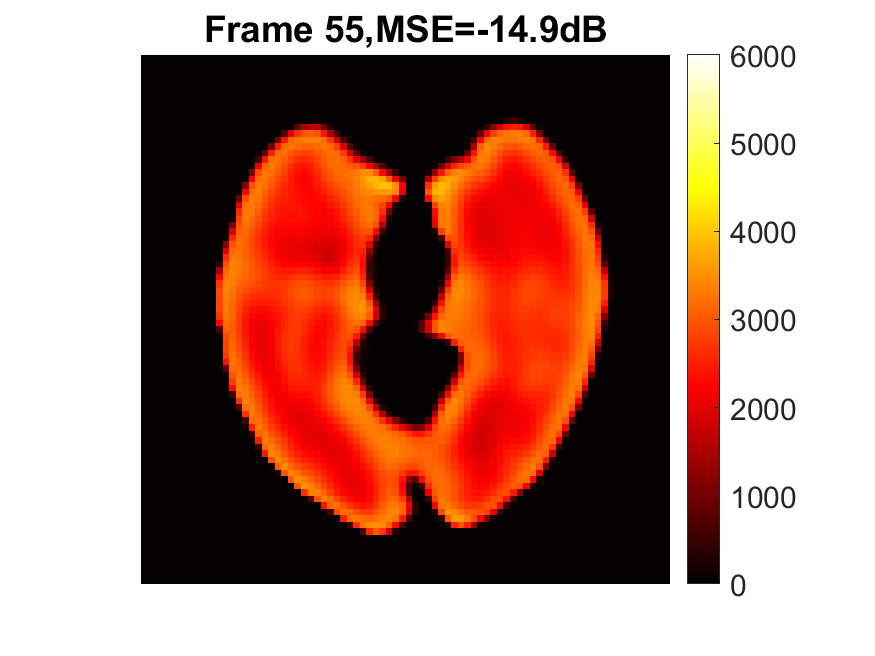}\\
			\end{minipage}%
		}%
		\subfloat[]{
			\begin{minipage}[t]{0.18\linewidth}
				\centering
				\includegraphics[trim=2cm 1cm 3.4cm 0cm, clip, height=3.2cm]{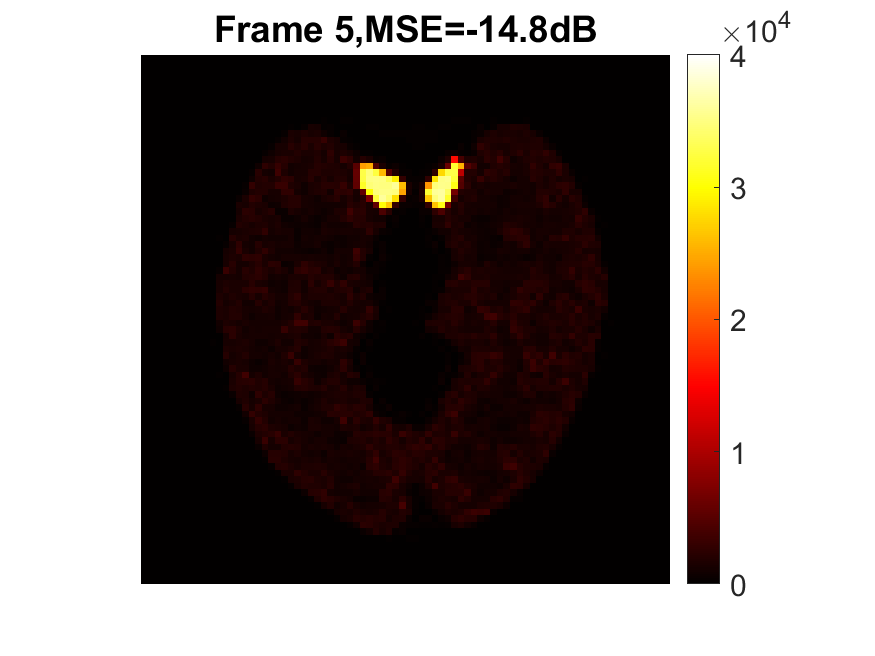}\\
				\includegraphics[trim=2cm 1cm 3.4cm 0cm, clip, height=3.2cm]{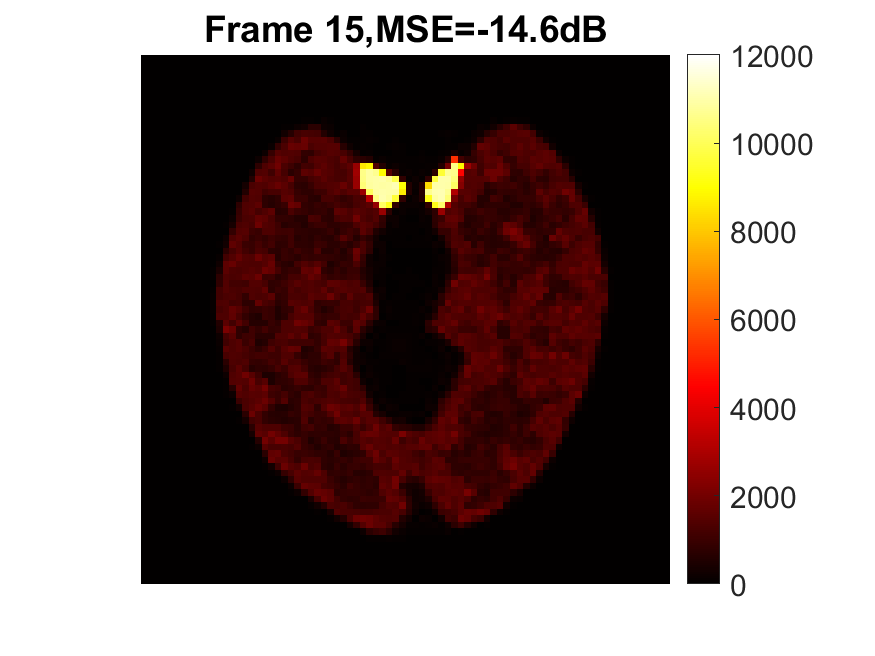}\\
				\includegraphics[trim=2cm 1cm 3.4cm 0cm, clip, height=3.2cm]{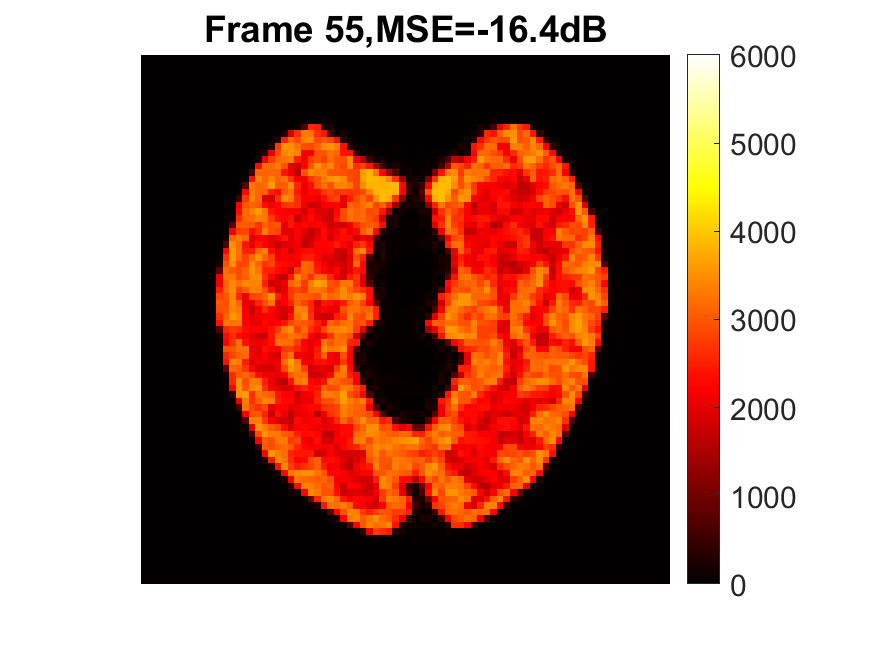}\\
			\end{minipage}%
		}%
		\subfloat[]{
			\begin{minipage}[t]{0.18\linewidth}
				\centering
				\includegraphics[trim=2cm 1cm 3.4cm 0cm, clip, height=3.2cm]{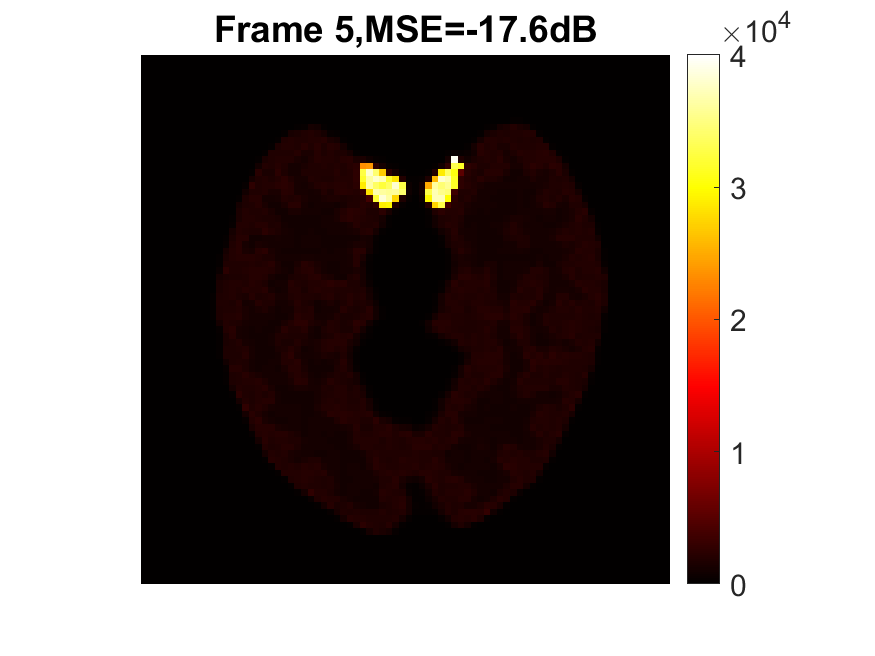}\\
				\includegraphics[trim=2cm 1cm 3.4cm 0cm, clip, height=3.2cm]{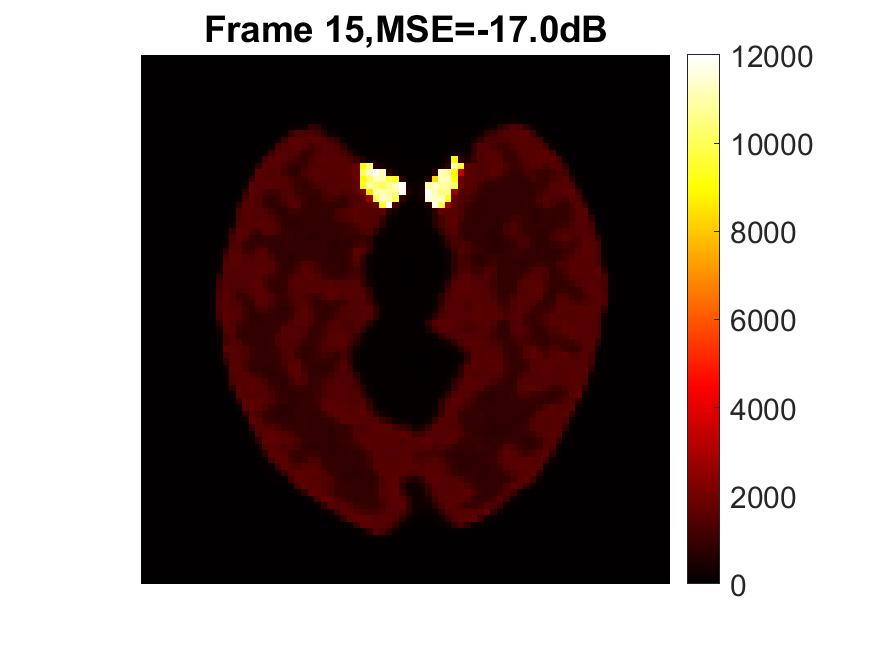}\\
				\includegraphics[trim=2cm 1cm 3.4cm 0cm, clip, height=3.2cm]{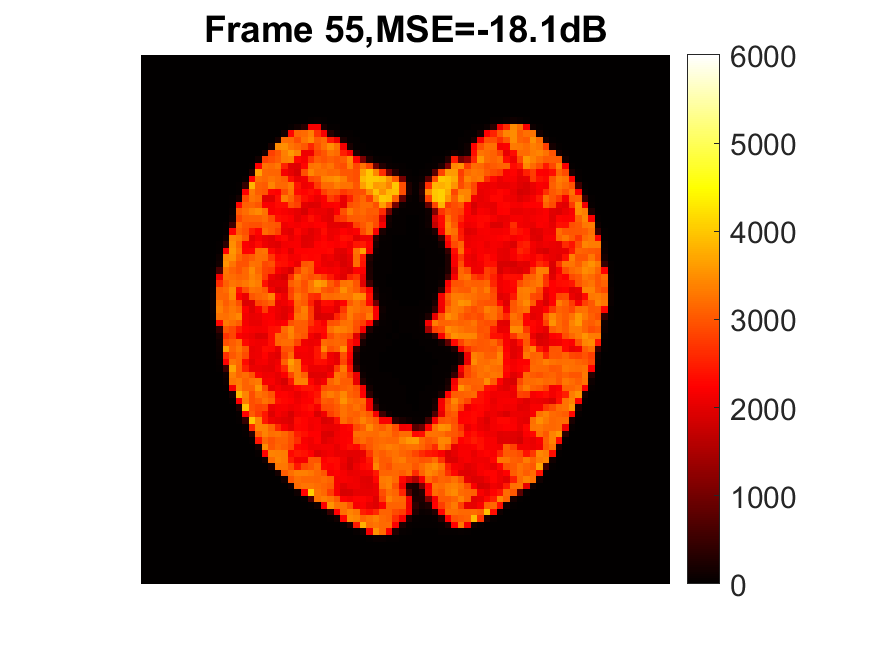}\\
			\end{minipage}%
		}%
		\caption{True activity images and reconstructed images by different methods for frame 5 (top row), frame 15 (middle row) and frame 55 (bottom row). (a) True images, (b) ML-EM, (c) DIP method, (d) conventional kernel method, and (e) proposed deep kernel method.}
		\label{fig:Rec}
	\end{figure*}
	\begin{figure*}[h!]
		\vspace{-20pt}
		\centering
		\subfloat[]{\includegraphics[trim=0cm 0cm 0.5cm 0cm, clip,width=2.5in]{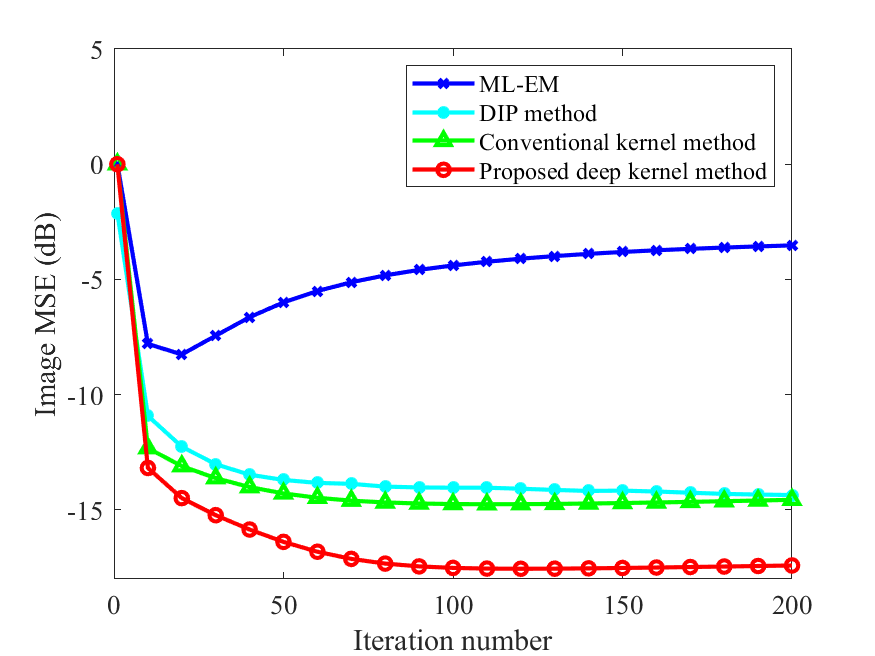}
			\label{fig_3_case}}
		\subfloat[]{\includegraphics[trim=0cm 0cm 0.5cm 0cm, clip, width=2.5in]{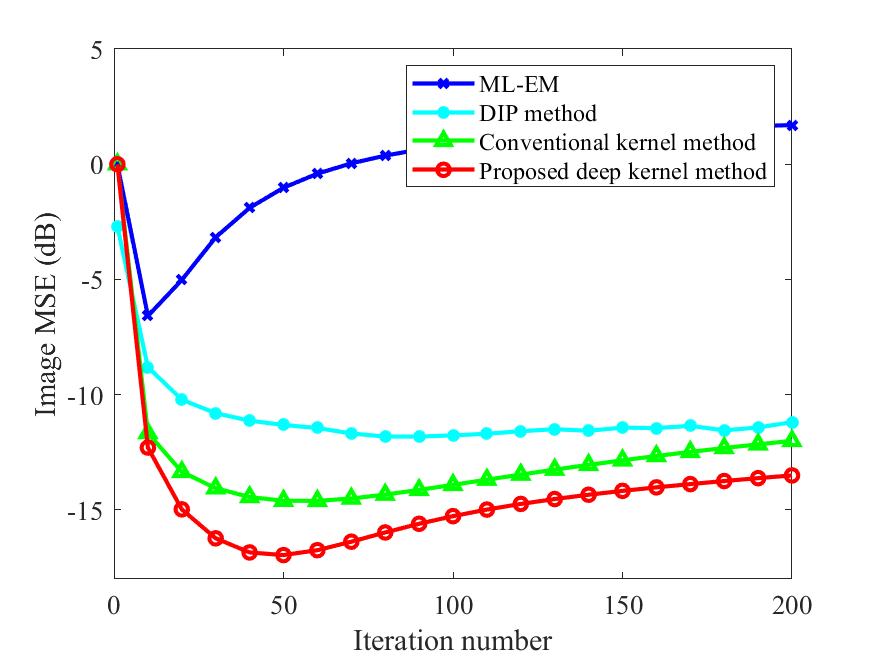}
			\label{fig_4_case}}\\
		\subfloat[]{\includegraphics[trim=0cm 0cm 0.5cm 0cm, 
			clip, width=2.5in]{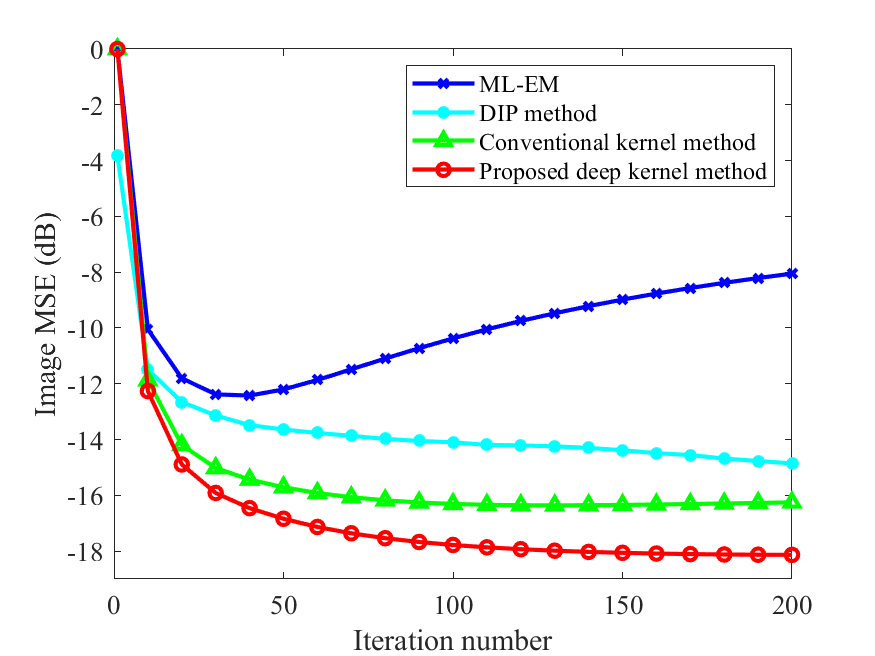}
			\label{fig_5_case}}
		\subfloat[]{\includegraphics[trim=0cm 0cm 0.5cm 0cm, width=2.5in]{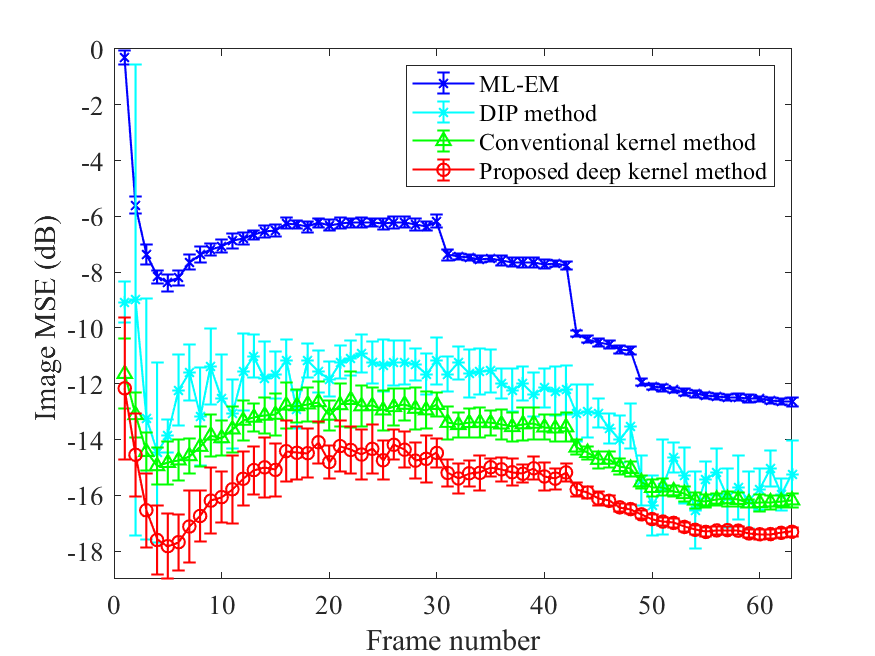}}
		\caption{Comparison of image MSE for different reconstruction methods. (a-c) plot of image MSE as a function of iteration number for (a) frame 5, (b) frame 15, and (c) frame 55; (d) image MSE of all time frames. The error bars in (d) were obtained from 20 realizations \txtb{and here the MSE of each frame was minimized over the iteration numbers in different methods.}}
		\label{fig:MSE}	
	\end{figure*}
	\begin{figure*}[t]
		\vspace{-30pt}
		\centering
		\subfloat[]{\includegraphics[trim=0.5cm 0cm 1cm 0cm, clip,width=2.2in]{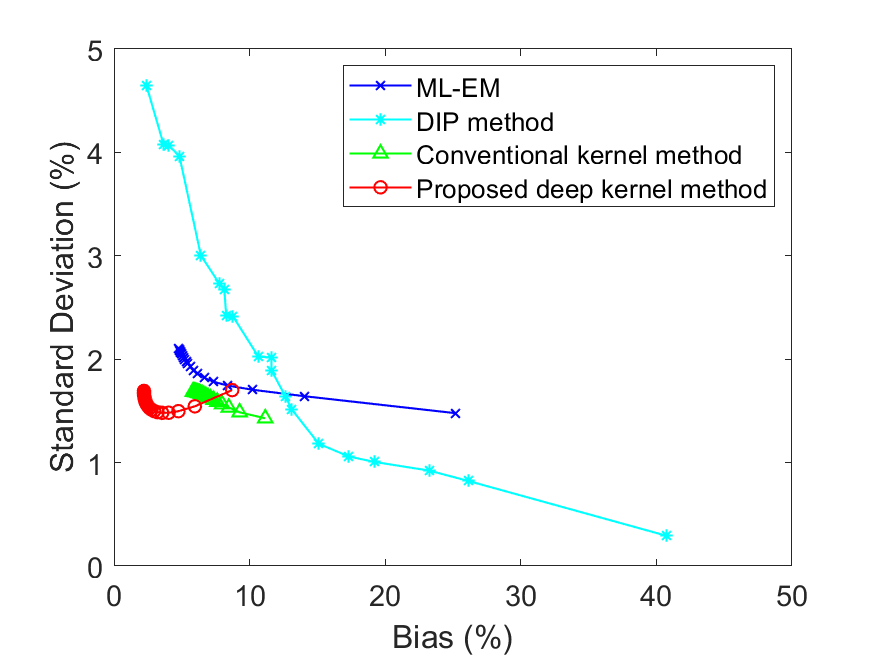}
			\label{fig_3_case}}
		\hfil
		\subfloat[]{\includegraphics[trim=0.5cm 0cm 1cm 0cm, clip, width=2.2in]{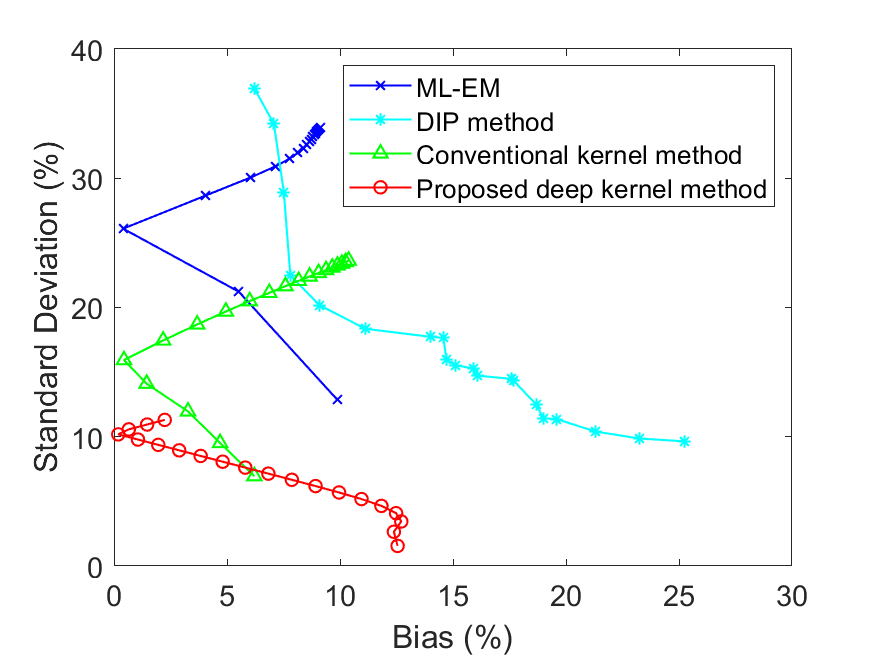}
			\label{fig_4_case}}
		\hfil
		\subfloat[]{\includegraphics[trim=0.5cm 0cm 1cm 0cm, 
			clip, width=2.2in]{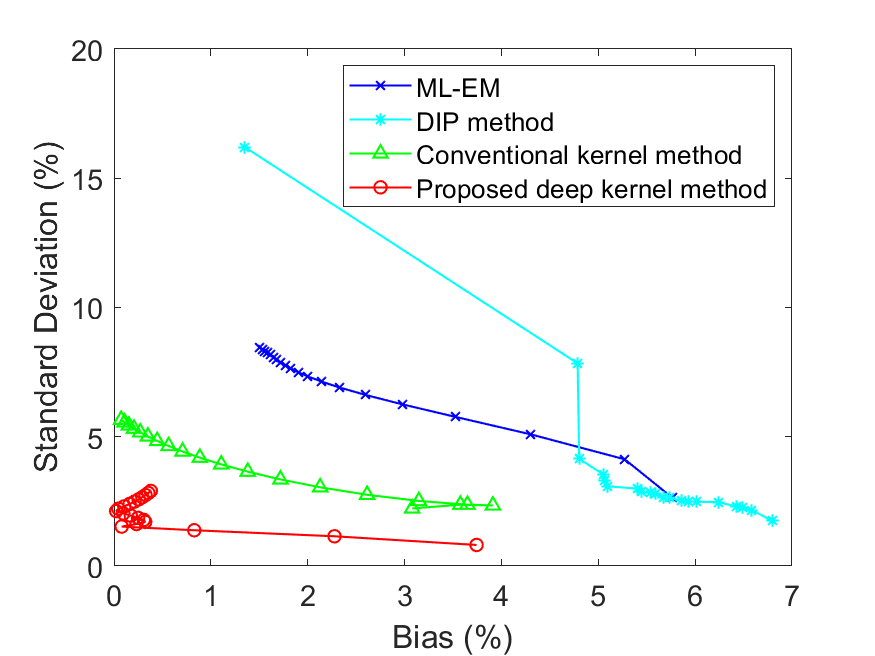}
			\label{fig_5_case}}
		\caption{Plots of bias-SD trade-off for ROI quantification by varying the iteration number from 10 to 200 with 10 intervals (i.e., from rightmost to leftmost on each curve). (a) Blood ROI in frame 5, (b) tumor ROI in frame 5, (c) tumor ROI in frame 55.}
		\label{ROI}
	\end{figure*}
	
	\subsection{Demonstration of Attention Map for the Kernel Methods}
	
	To understand how the deep kernel method may improve image reconstruction, Fig. \ref{fig:attention} illustrates the attention maps for two different ``query'' pixels, one from the tumor and the other from the white matter region. These attention maps were generated by reshaping the corresponding row of the kernel matrix for a query pixel $j$. 
	The traditional ML-EM reconstruction can be considered as a special pixel-kernel method for which the kernel matrix is the identity matrix. As illustrated in Fig. \ref{fig:attention}(b), the attention of ML-EM just focuses on the query pixel $j$ itself. No spatial correlation is explored by this pixel kernel.
	
	The conventional kernel method \cite{Wang2015} is able to exploit spatial correlation from pixels that are considered as neighbors of the query pixel $j$ by kNN. The attention is not only on the query pixel but also spreads nonlocally to neighboring pixels (``key'') in the whole image. However, these ``key'' pixels may be falsely identified if $k$ in kNN is large (here $k=200$) \cite{Wang2015}. Without deep learning, the existing kernel model is unable to exclude the effect of those false neighbors. For example, as shown in Fig. \ref{fig:attention}(c), ``key'' pixels in the gray-matter and white-matter regions were falsely assigned with high attention for a query pixel from the tumor, and ``key'' pixels in the gray-matter were also falsely assigned with high attention for a query pixel from the white-matter region. 
	
	In comparison, the deep kernel model with training can \txtb{learn feature extraction from data, which leads to} a more appropriate weight to irrelevant ``key'' pixels even if those pixels are initially included in the $k$ nearest neighbors. Fig. \ref{fig:attention}(d) shows that with deep learning, attention is predominantly extracted in the tumor region for the tumor query pixel and in the white-matter region for the white-matter query pixel.
	
	\subsection{Image Quality Comparison}
	Fig. \ref{fig:Rec} shows the ground-truth activity images and reconstructed images by different reconstruction methods for frame 5 (an early 2-s frame, low count level), frame 15 (a middle 2-s frame, low count level) and frame 55 (a late 1-min frame, relatively high count level), respectively. The results of image MSE in dB are included. The kernel-based methods ((d) and (e)) both achieved a better image quality with lower MSE as compared to the methods without kernel ((b) and (c)). The DIP method \cite{Gong2019} suppressed noise well but also resulted in
	over-smoothness. The proposed deep kernel method achieved a better image quality with lower MSE as compared to other three methods thanks to the \txtb{improved} attention weights embedded in the learned kernel matrix $\K$.
	
	Fig. \ref{fig:MSE}(a-c) further show the image MSE plots of frame 5, frame 15 and frame 55 by varying the iteration number in each reconstruction algorithm. For the DIP reconstruction, the results of the first iteration were always better due to \txtb{the use of four sub-iterations} for the tomographic reconstruction step in the ADMM algorithm. The proposed deep kernel method demonstrated a substantial improvement at all later iterations over the conventional kernel method and the DIP method.
	
	The MSE results of all time frames are shown in Fig. \ref{fig:MSE}(d). Here shown are the best MSE (over different iterations) for each frame in different methods. Error bars were calculated over 20 noisy realizations. 
	The DIP method showed an unstable behavior across different frames. In contrast, the deep kernel demonstrated a significant improvement over other methods.
	
	\txtb{Note that image MSE is only an indicator of global image quality and does not reflect task-specific evaluation. Its weakness is compensated by the ROI quantification results presented in the next subsection.}
	
	\subsection{ROI Quantification Comparison}
	Fig. \ref{ROI} shows the trade-off between the \txtb{absolute} bias and SD of different methods for ROI quantification in a blood ROI
	(Fig. \ref{ROI}a) and a tumor ROI (Fig. \ref{ROI}b and Fig. \ref{ROI}c). The curves were obtained by varying the iteration number from 10 to 200 iterations with an interval of 10 iterations. Note that the uptake in the blood region reached its maximum in frame 5 as shown in Fig. \ref{fig:phant}b. At a comparable bias level, the proposed deep kernel had a lower noise SD than the conventional kernel method for both the blood ROI and tumor ROI. The results by the DIP method were even worse than the ML-EM results due to over-smoothness, though it had a better image MSE performance as shown in Fig. \ref{fig:MSE}. \txtb{Some curves in Fig. \ref{ROI}b  show a sharp change of direction because the bias at early iterations was negative and became positive due to high noise at late iterations.} 
	
	\begin{figure}[t]
		\centering
		\includegraphics[trim=1cm 0cm 1cm 0cm, width=2.5in]{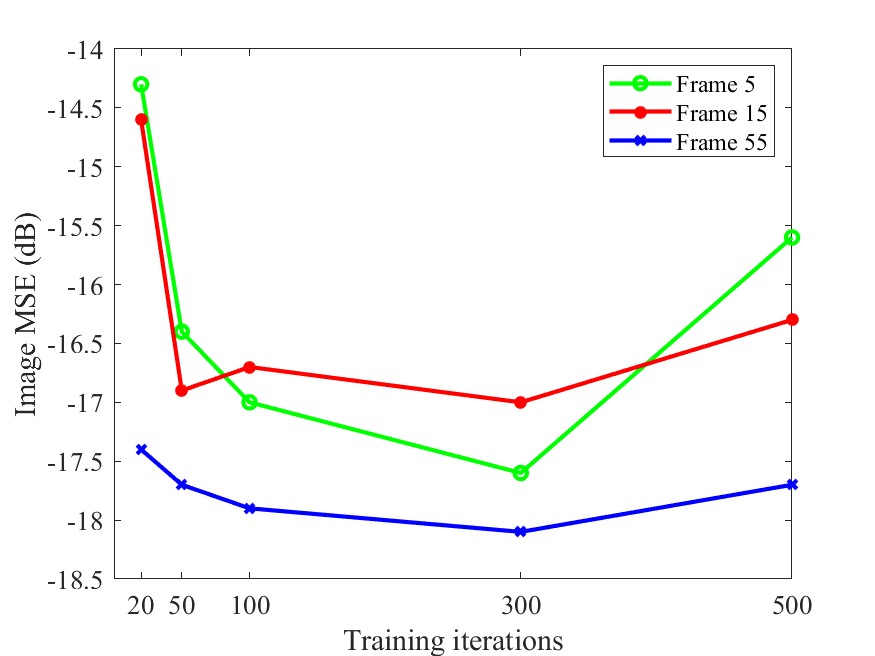}
		\caption{Effect of training iterations on the MSE performance of the proposed deep kernel for three different time frames.}
		\label{fig:training_iterations}	
	\end{figure}
	
	
	\begin{figure*}[t]
		\centering
		\subfloat[Frame 19 at $t = 36-38s$]{\includegraphics[trim=0cm 0cm 0cm 0cm, clip,width=7in]{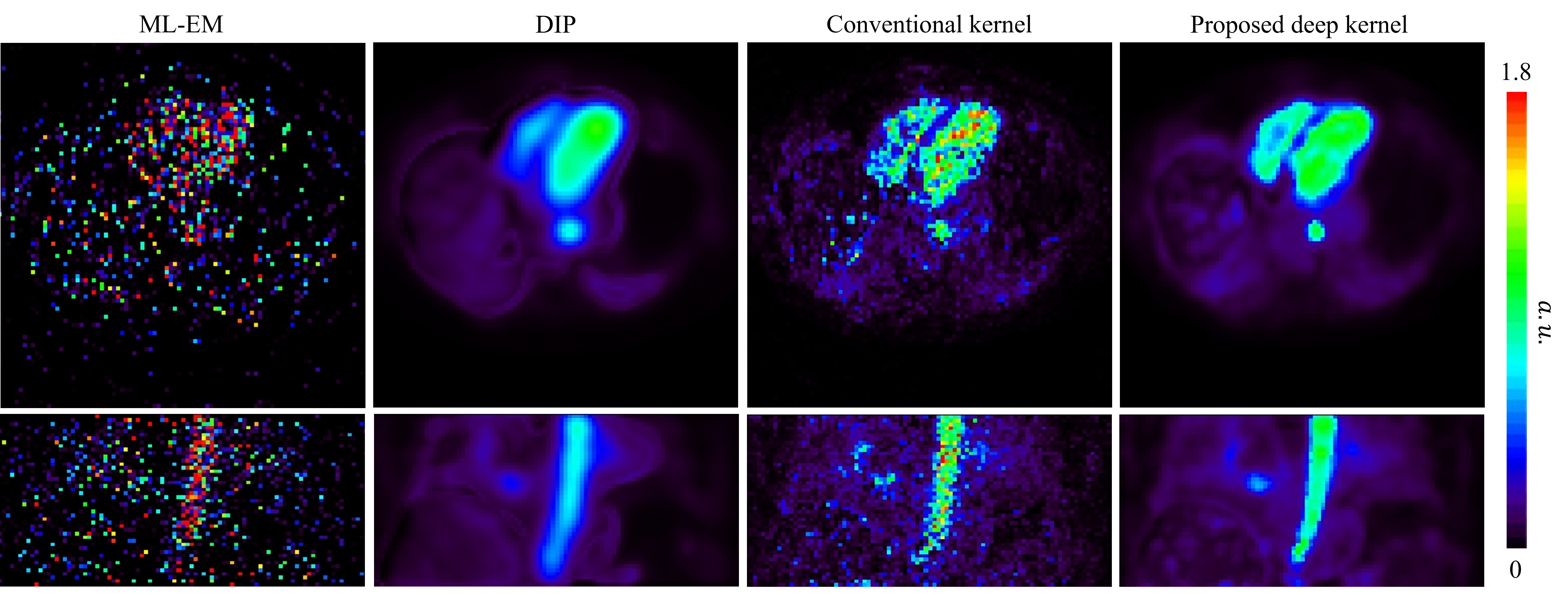}
			\label{fig_1_case}}
		\hfil
		\subfloat[Frame 73 at $t = 144-146s$]{\includegraphics[trim=0cm 0cm 0cm 0cm, clip, width=7in]{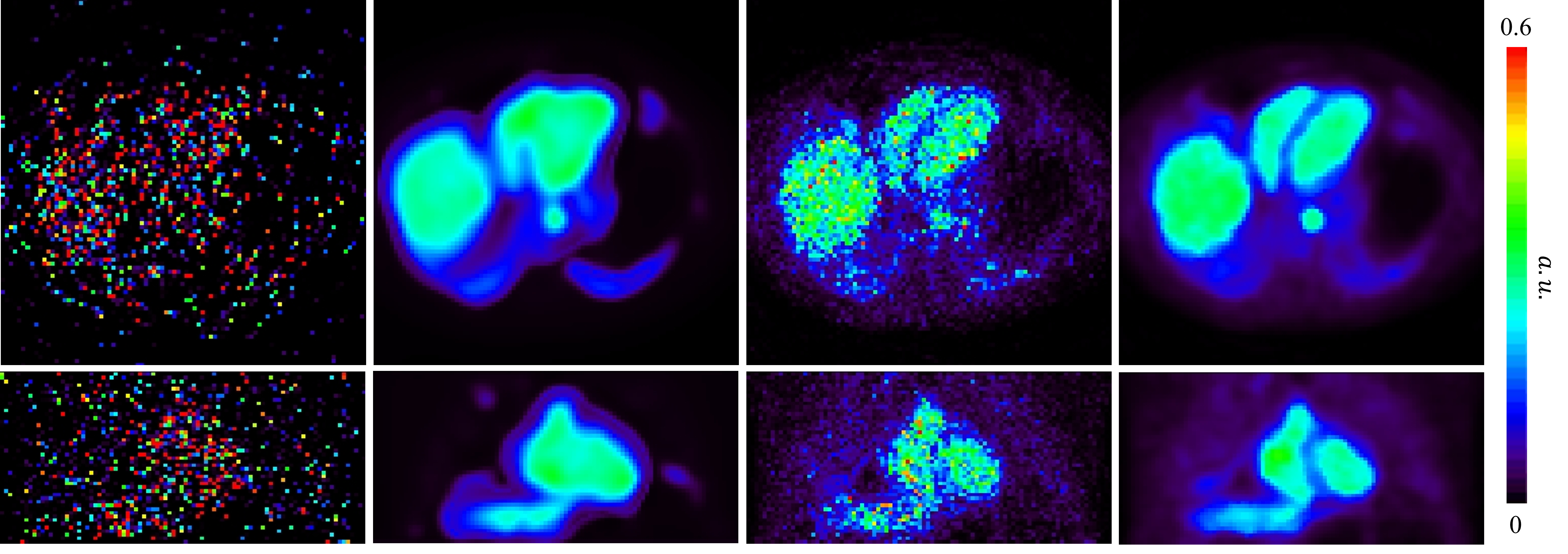}
			\label{fig_2_case}}
		\caption{Reconstruction of high-temporal resolution frames (2s/frame) at (a) $t=36-38s$ and (b) $t=144-146s$ by different methods: ML-EM, DIP method, conventional kernel method and proposed deep kernel method.}
		\label{fig:MV}
	\end{figure*} 
	
	\subsection{Effect of Method Parameters}
	
	One parameter that has an important effect on the deep kernel method is the number of training \txtb{iterations}. With increasing iteration number, the training loss was steadily reduced but the corresponding final performance of the PET reconstruction results did not follow this trend. Fig. \ref{fig:training_iterations} shows the effect of training iterations on the MSE performance for frame 5, frame 15, and frame 55. The quality of reconstructed PET images may become worse if the training iteration number is too large. This is because the training may start to fit the noise in the composite image prior and the resulting error can be propagated into the trained kernel matrix and final reconstruction. The result here suggests a reasonable choice was 300 iterations, which also worked well for all other frames.
	
	\section{Application to Patient Data}
	
	\subsection{Data Acquisition}
	
	A cardiac patient scan was performed on the GE Discovery ST PET/CT scanner in 2D mode at the UC Davis Medical Center. The patient received approximately 20 mCi $^{18}$F-FDG with a bolus injection, followed by an immediate dynamic scan. The one-hour data are divided into 109 time frames following the schedule $75 \times 2s$, $15 \times 10s$, $10 \times 60s$, and $9 \times 300s$. A low-dose transmission CT scan was performed at the end of PET scan for PET attenuation correction. The projection data size was $249\times210\times47$ and the image size was $128\times128\times47$ with a voxel size of $3.91\times3.91\times3.27$ mm$^3$. The data correction sinograms of each frame, including normalization, attenuation correction, scattered correction and randoms correction, were extracted using the vendor software and used in the reconstruction process. 
	
	\begin{figure*}[t]
		\vspace{-0pt}
		\centering
		{\includegraphics[trim=0cm 0cm 0cm 0cm, clip,width=7in]{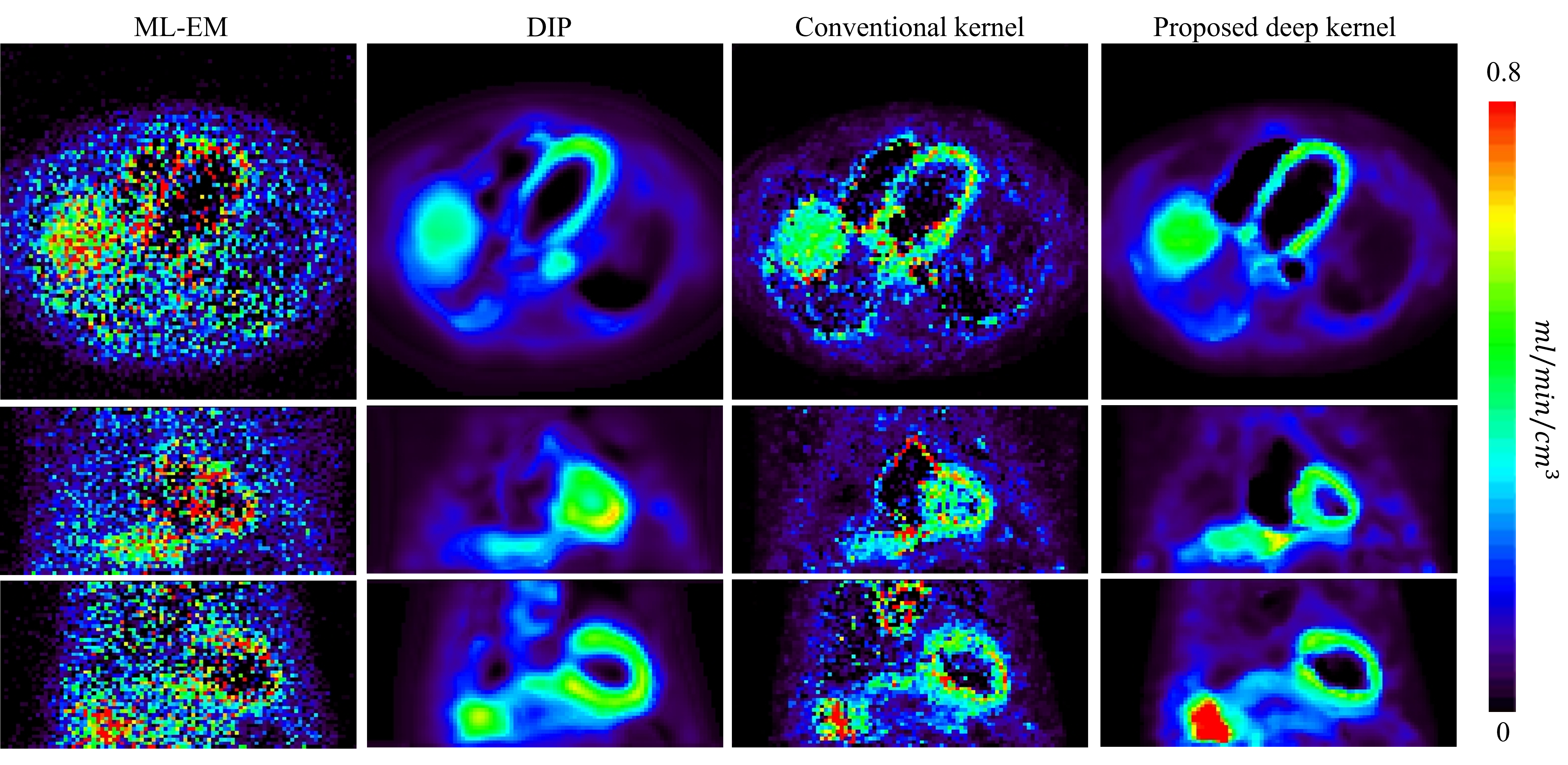}
			\label{fig_1_case}}
		\caption{Parametric images of $K_1$ generated from the early-dynamic images reconstructed using ML-EM, DIP method,  conventional kernel method and proposed deep kernel method. Each image is shown in transverse, coronal and sagittal views.}
		\label{K1 image}
	\end{figure*} 
	\begin{figure}[h]
		\vspace{-0pt}
		\centering
		{\includegraphics[trim=0cm 0cm 1cm 0cm, clip,width=3in]{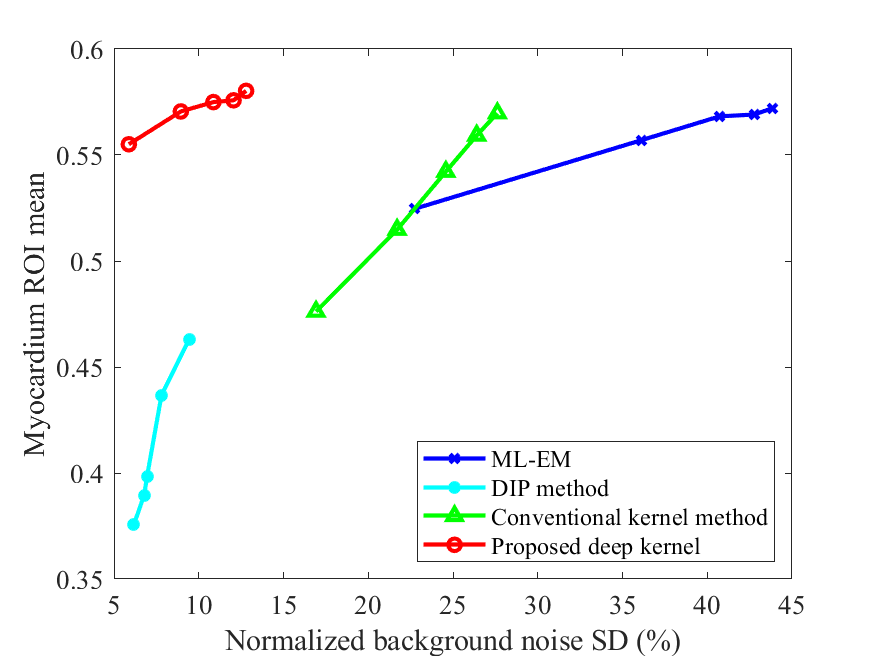}
			\label{fig_1_case}}
		\caption{Plots of ROI mean of myocardial $K_1$ versus liver background noise by varying the reconstruction iteration number from 20 to 100 in each method.}
		\label{CRC_K1}
	\end{figure}
	\subsection{Results of Reconstructed PET images}
	We compared the proposed deep kernel method with the ML-EM, DIP method\cite{Gong2019} and conventional kernel method \cite{Wang2015}. Because the ADMM algorithm resulted in a very poor DIP reconstruction for this patient dataset, here we instead used the optimization transfer algorithm \cite{Li2022} for the DIP method. Details and advantages of the OT algorithm are \txtb{described in}  \cite{Li2022}. The prior images used in the two kernel methods and the DIP method were obtained using four composite images that were reconstructed from four composite frames (one 5-min frame, one 15-min frame and two 20-min frames).  Other implementation settings were as the same as the simulation study. The $k$ in kNN for defining the neighborhood {$\mathcal{N}_j$} was also set to be 200. All the methods were run for 100 iterations starting from a uniform initial image. 
	
	Fig. \ref{fig:MV} shows the reconstructed activity images using different algorithms for two early-time high-temporal resolution (HTR) frames (2s/frame), one at $t=37$s and the other at $t=145$s. The ML-EM reconstructions were extremely noisy due to the low-count level. The conventional kernel method led to substantial noise reduction but additional noise still remained. Similar to the simulation results, the DIP method successfully suppressed the high noise but also resulted in oversmoothed images and inconclusive separation between the left ventricle and right ventricle. In comparison, the images by the proposed deep kernel method demonstrated a significant improvement with clearer structures and lower noise in the left ventricle cavity and myocardium, though no ground truth is available for the real dataset.
	
	
	\subsection{Demonstration for Parametric Imaging}
	
	Parametric imaging was also performed for the dynamic
	images of the same subject using a two-tissue compartment model \cite{Zuo2020}. \txtb{We used the classic Levenberg–Marquardt algorithm with 50 iterations to solve the optimization problem and the fitting process was implemented using c/c++ programing \cite{Wang2012}.} For each method, the left ventricle region was used to extract an image-derived input function. Because different reconstruction methods mainly make a difference for early-time frames which have a low count level (Fig. \ref{fig:MV}), here we focused on parametric imaging of early-dynamic data using the first 150 seconds.
	
	Fig. \ref{K1 image} shows the parametric images of FDG delivery rate $K_1$. The ML-EM result suffered from heavy noise. The conventional kernel method demonstrated an improvement but still suffered from noise and artifacts. The DIP method largely reduced the noise but also resulted in oversmoothness. It also led to a high  $K_1$ value in the aorta region compared to other three methods. In comparison, the $K_1$ image obtained by the proposed deep kernel method substantially suppressed the noise and showed a more continuous and clearer myocardium.
	
	Fig. \ref{CRC_K1} further shows a quantitative comparison of different methods for myocardial ROI quantification in the $K_1$ image. Here the ROI mean is plotted versus normalized background noise SD by varying the iteration number from 20 to 100 with an interval of 20 iterations. The conventional kernel method outperformed the ML-EM reconstruction noticeably. For a given myocardial $K_1$ value, the conventional kernel method had a lower liver background noise SD than ML-EM. The DIP method resulted in underestimation of myocardial $K_1$ compared to other three methods, though the noise was suppressed well. The proposed deep kernel method achieved a better trade-off than all other three methods. For a given ROI mean value (e.g., 0.57), for example, the deep kernel method had the lowest background noise level as compared to the ML-EM and conventional kernel methods. The deep kernel method also had a higher myocardial $K_1$ value than the DIP method for a given noise level (e.g., 8\%) in the liver background.  \txtb{The higher $K_1$ value was closer to the myocardial ROI mean quantified with the ML-EM reconstruction.}
	
	\section{Discussions}
	
	This paper proposed a deep kernel method that learns the trainable components of the neural network model from image prior to enable automated learning of \txtb{an improved} kernel method. Compared to the conventional kernel method \cite{Wang2015} that builds the kernel representation using an empirical process, the proposed deep kernel method can \txtb{learn to extract a more appropriate feature set for building improved kernels} from the data, as illustrated in Fig. \ref{fig:attention}. Compared to the DIP method \cite{Gong2019} that introduces a complex non-linear learning in the reconstruction, the deep kernel method only introduces the non-linear learning into the kernel representation, but remains a linear representation for the kernel coefficient image and is therefore easy to be reconstructed from PET data. The comparison results from the simulation and real data studies indicate a better performance of the deep kernel method than other methods. 
	
	\txtb{Similar to the conventional kernel method \cite{Wang2015} and the DIP method \cite{Gong2019}, the proposed method is directly applicable to single subjects, which has been demonstrated for dynamic PET in this paper but can be potentially extended to static image reconstruction if a training pair becomes possible. The prior image must be of relatively high quality. When noise presents, early-stopping can be used to avoid overfitting in the training. Alternatively, regularized training may be explored to address the challenge.}
	
	The deep kernel method in this work focused on frame-by-frame image reconstruction in the spatial domain but can be potentially extended to the spatiotemporal domain as used in \cite{Wang2019}. The kernel coefficient image $\alp$ in the deep kernel model can be also further parameterized using a neural network, in a way similar to our \txtb{other} work \cite{Li2022}. \txtb{In addition, the current study only used the kernel form following the Gaussian function and Euclidean distance. However, it is possible to train an optimized kernel form from the prior data. } These modified but more complex methods will be explored in our future work.
	
	Compared to the standard kernel method, the learning of a deep kernel adds an extra computational cost. For the 3D real data study, the training time was 20 minutes as compared to half a minute for the construction of a conventional kernel matrix. However, the extra computational cost may be relatively small when compared to the time (ranging from 30 minutes to several hours) required for the actual \txtb{kernelized EM} reconstruction step (see (\ref{KEM})) for a dynamic PET scan. In addition, the extra time can be further reduced if a large database becomes available to pre-train the optimal kernel construction, for example, using high performance total-body PET scanners (e.g., \cite{Cherry2017, Badawi2019, Spencer2021, Karp2018, Pantel2020}), which will also be explored in our future work.
	
	\section{Conclusion}
	
	In this paper, we have developed a new deep kernel method for PET image reconstruction. The proposed deep kernel model allows the construction of kernel representation to be trained from data rather than defined by an empirical process. Computer simulation and patient results have demonstrated the improvement of the deep kernel method over existing methods in dynamic PET imaging. 
	
	\section*{Acknowledgment}
	
	\txtb{The authors thank Dr. Benjamin Spencer, Dr. Yang Zuo, and Mr. Michael Rusnak for their assistance in patient data collection.}

\end{document}